\documentclass[conference]{IEEEtran}
\usepackage{epsfig}
\usepackage {graphicx} 
\usepackage {amsthm}
\usepackage[fleqn] {amsmath}
\usepackage {amsfonts, amssymb}
\usepackage{multirow}

\usepackage{enumerate}
\usepackage{color}
\usepackage{algorithmic} 
\usepackage[]{algorithm}
\usepackage{graphicx}
\usepackage{caption}
\usepackage{subcaption}
\usepackage{datetime}

\usepackage{url}

\usepackage{flushend}

\begin{document}

\title{Large-Scale Distributed Internet-based Discovery Mechanism for Dynamic Spectrum Allocation}

\author{
    
\IEEEauthorblockN{
    Magnus Skjegstad\IEEEauthorrefmark{1}\IEEEauthorrefmark{2}\IEEEauthorrefmark{4},
    Brage Ellings{\ae}ter\IEEEauthorrefmark{1}\IEEEauthorrefmark{2}\IEEEauthorrefmark{4},
    Torleiv Maseng\IEEEauthorrefmark{1}\IEEEauthorrefmark{2}\IEEEauthorrefmark{4}, 
    Jon Crowcroft\IEEEauthorrefmark{5} and
    {\O}ivind Kure\IEEEauthorrefmark{2}\IEEEauthorrefmark{3}}

\IEEEauthorblockA{
    \IEEEauthorrefmark{1}University of Oslo}

\IEEEauthorblockA{
    \IEEEauthorrefmark{2}University Graduate Center, Kjeller (UNIK)}

\IEEEauthorblockA{
    \IEEEauthorrefmark{3}Norwegian University of Science and Technology (NTNU)}

\IEEEauthorblockA{
    \IEEEauthorrefmark{4}Norwegian Defence Research Establishment (FFI)}

\IEEEauthorblockA{
    \IEEEauthorrefmark{5}University of Cambridge}
}
    
\maketitle

\begin{abstract}
Scarcity of frequencies and the demand for more bandwidth is likely to increase the need for devices that utilize the available frequencies more efficiently. Radios must be able to dynamically find other users of the frequency bands and adapt so that they are not interfered, even if they use different radio protocols. As transmitters far away may cause as much interference as a transmitter located nearby, this mechanism can not be based on location alone. Central databases can be used for this purpose, but require expensive infrastructure and planning to scale. In this paper, we propose a decentralized protocol and architecture for discovering radio devices over the Internet. The protocol has low resource requirements, making it suitable for implementation on limited platforms. We evaluate the protocol through simulation in network topologies with up to 2.3 million nodes, including topologies generated from population patterns in Norway. The protocol has also been implemented as proof-of-concept in real Wi-Fi routers.
\end{abstract}

\section{Introduction}

Many radio transmitters we use daily are connected to the Internet. Wi-Fi routers are a typical example, but recently other devices, such as femtocells have appeared. The scarcity of frequencies combined with the demand for more bandwidth is likely to further increase the need for devices that provide high wireless bandwidth locally while using a wired network to carry data over longer distances. Locally available frequencies can be utilized more efficiently by enabling these devices to coordinate with other radio devices which could be interfered or interfere in their area.

Cognitive Radio (CR) and Dynamic Spectrum Access (DSA) are technologies that can help alleviate the coming spectrum shortage. So far, most research has been focused on physical layer and MAC layer capabilities of such systems, and recently some standards (such as IEEE 802.22 \cite{wran}) have emerged. When investigating physical layer performance of CR and DSA networks one usually assumes a network consisting of 10-50 nodes. However, a network of radio devices connected to the Internet will consist of thousands to millions of nodes distributed over large areas, even countries. With this vast amount of different radios it is important to be able to find other radios to communicate with and also radios one needs to coordinate traffic with. The design space for such a solution range from a fully centralized allocation to a fully distributed one without any centralized control. This paper focuses on the feasibility of the latter, proposing a system design that can be used on a country/community scale. 

The issue of centralized and decentralized is not only a technical issue, but also a political one concerning the control over frequency resources. FCC has proposed using a database for discovering available frequencies in the US\@. The database will contain areas where it is safe to use radio transmission in part of the white space TV frequencies. The system is dimensioned to take care of the TV viewers without knowing their location by making worst case assumptions. How the database is to be accessed is about to be defined by the Protocol to Access White Space database~\cite{rfc6953}. However, this system is currently not designed to allow fine grained discovery.

A disadvantage of a centralized scheme is that it requires an organization willing to take responsibility and the cost associated with establishing and maintaining the infrastructure. For some frequency bands, the owner of the frequencies is an obvious candidate. In other bands, such as the open ISM band, there may be no clear candidates for taking this responsibility. An alternative design option is a fully decentralized system similar to existing peer-to-peer (P2P) solutions. They can grow organically as the user numbers and requirements grow. As long as there is a community of users with the same interest they will function. There is no need for a large initial investment or maintenance. Each user carries their own cost and the community ensures development and maintenance of the software and system.  

A P2P client could use existing network connections to build an overlay over the Internet and then use it to discover and negotiate with other radio nodes in their area. This requires a connection to the Internet, which is also a requirement of a centralized solution. As the client can run directly on the radio node, the system would not require additional investment in infrastructure or central servers. The decentralized solution may be gradually deployed and coexist with legacy systems and the performance would increase as more devices added support for it. A decentralized protocol may even be used in addition to central databases, e.g., by being used as a fall-back mechanism in areas that are not covered by other solutions or to discover databases that are responsible for frequency allocation in a given area.

The Norwegian Post and Telecommunications Authority (NPT) is concerned about effective usage of frequency bands exclusively used by primary users. In their 2012-2014 strategy~\cite{nptstrategy}, NPT argues that DSA and CR should be used to allow secondary users as long as they maintain knowledge about other users in their vicinity and do not disturb the primary users. NPT states that they will work to include this requirement in new spectrum licenses. Secondary users are also permitted in spectrum licenses awarded by the Swedish Post and Telecom Authority (\cite{pts1} and \cite{pts2}).

When many secondary users use the same spectrum they need to be able to coordinate their access with both the primary user and each other. However, the secondary users may use different radio protocols and therefore not be able to communicate directly by radio. A potential use case for a decentralized discovery protocol is thus to enable secondary users connected to the Internet to coordinate their transmissions to satisfy regulatory conditions and maximize their joint link performance.

In this paper, we propose and implement a decentralized P2P protocol for large-scale discovery of radio devices over the Internet. The protocol has low bandwidth-, memory- and processing requirements, making it suitable for running on platforms with limited resources, such as future Wi-Fi routers or femtocells. We evaluate the protocol through simulation in Internet-scale network topologies, including topologies generated from population patterns in four municipalities in Norway. Finally, we propose a generic architecture for development of new discovery mechanisms and resource allocation algorithms in large hybrid- or fully distributed systems. 

The contributions of this paper are

\begin{itemize}
    \item A large-scale, decentralized discovery protocol for discovering radio devices over the Internet
    \item A generic architecture that provides a clear separation of concern for radio device discovery and resource allocation
    \item An extensive evaluation of the proposed protocol in large, simulated networks
\end{itemize}

Some of the components described in this paper are adapted from approaches published on their own in earlier works. The key contribution of this paper is thus the combination of these techniques to implement a large scale, decentralized system for frequency allocation. To the best of our knowledge, this is the first proposed protocol for decentralized DSA on this scale.

This paper is organized as follows. In Section~\ref{sec:arch} we propose a layered architecture for frequency allocation and discovery. In Section~\ref{sec:coord} we propose a simplified propagation model for discovering radio nodes and proceed to present the discovery protocol in Section~\ref{sec:disco}. The protocol is evaluated in Section~\ref{sec:eval}. Related work is presented in Section~\ref{sec:relwork} and Section~\ref{sec:conclusion} concludes the paper.
 
\section{Architecture}
\label{sec:arch}

\begin{figure}
\centering
\includegraphics[width=0.95\columnwidth]{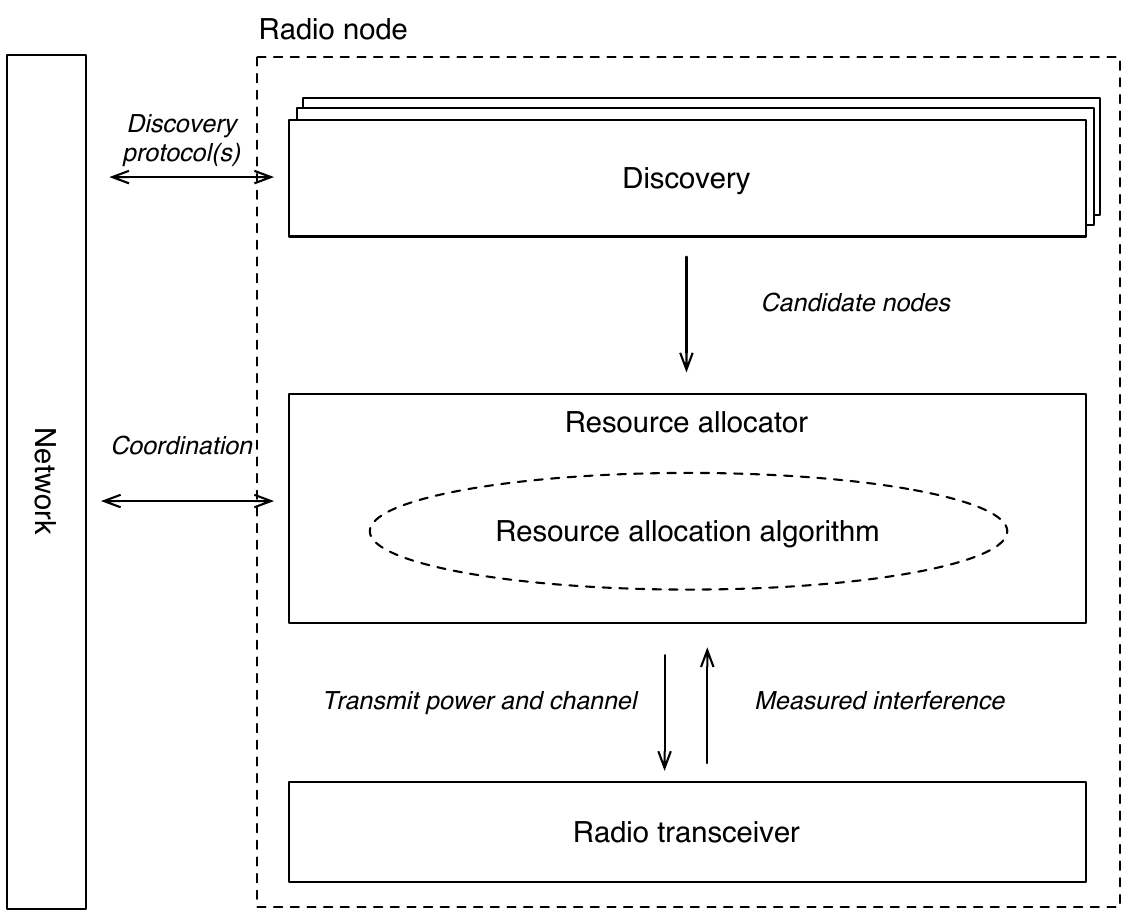}
\caption{Discovery architecture overview.}
\label{fig:p2p_allocator}
\end{figure}

A discovery protocol is an auxiliary mechanism that provides information to other components in a larger architecture. As the design of the architecture influences the design and requirements of the discovery protocol, we start by describing the envisioned architecture in more detail. We argue that in large, heterogeneous networks, it is useful to separate frequency allocation from the discovery process, as this provides a clear separation of concern, as well as a well-known information platform that can be reused between different algorithms. The architecture proposed here is generic to radio devices connected to the Internet and is not specific to a decentralized system. However, we mainly consider the case where \emph{all tasks are performed by the node itself}.

We assume that each radio node requires three main tasks to be performed: a) gather information about other nodes, b) perform resource allocation and negotiation based on this knowledge and c) configure its hardware to use the allocated resources. 

Figure~\ref{fig:p2p_allocator} shows the relationship between the three tasks and the information that must be passed between them. The first task is \emph{discovery}. In this task, information is gathered about other nodes. The information can be retrieved with the help of a distributed protocol or a central database. The main goal of this task is to select a set of \emph{candidate nodes} that must be taken into account during the resource allocation. This is the task that is performed by the discovery protocol proposed in this paper. Note that several mechanisms can be used simultaneously and the results can be aggregated in the same result set. For example, results from a database could be combined with results from a P2P protocol. 

When a node has selected a set of candidate nodes, it must begin to allocate resources. This is handled by a separate abstraction we refer to as the \emph{resource allocator}. The main responsibility of the node's resource allocator is to execute a resource allocation algorithm and to provide support functions needed by the algorithm. It is up to the allocator to coordinate with and gather information from the candidate nodes after they have been identified. The gathered information could for example include detailed radio parameters of other nodes or sensing data. The specific communication protocol used by the resource allocator is outside the scope of this paper, but we assume that the allocator is able to contact other nodes directly via a network connection or via a system representing them, e.g., through a standardised configuration exchange protocol. The resource allocator is ultimately responsible for configuring the \emph{radio transceiver}, which is the final task. 

In this architecture, the candidate nodes are selected with limited knowledge about the underlying radio system. The discovery tasks are unable to accurately model propagation, interference and so on. This means that the set of candidate nodes is an inaccurate representation of the nodes that must be taken into account by the resource allocator. To maintain the separation between discovery and resource allocation, the discovery task is thus required to provide a \emph{sufficient set} of candidate nodes based on a known, simplified propagation model. It is then up to the resource allocator reduce this set.

The motivation for selecting an inaccurate set of nodes before performing resource allocation is to quickly be able to reduce a large set of nodes (potentially millions) to a small set of nodes that is manageable for a computationally demanding resource allocator. The discovery task can be seen as a coarse-grained node selection, while the resource allocator performs fine-grained selection (and coordination).

An example of such a division of tasks between the discovery mechanism and resource allocator can be illustrated with IEEE 802.22 (WRAN) for operation in the TV white spaces \cite{wran}. The discovery mechanism provides knowledge of nearby access points and their Internet network addresses to the resource allocator component. The resource allocator can then contact access points that use the 802.22 standard and exchange information required for this standard for spectrum sharing.

In the following we describe how a decentralized discovery mechanism can be implemented for this architecture. The responsibilities of the resource allocator are further discussed in Section~\ref{sec:ra_introduction}, but are not the main focus of this paper.

\section{Coordination area}
\label{sec:coord}

As we described in Section~\ref{sec:arch}, the Internet-based discovery protocol must be able to provide a worst case set of candidate nodes to the resource allocator based on a simplified propagation model. The model we use in this paper assumes a circular boundary around each node, which we refer to as the node's \emph{coordination area}. This is the estimated area in which the node may interfere with or be interfered by other radio transceivers. The goal is thus to discover the Internet address of all other nodes that have a coordination area that overlaps with the node running the discovery protocol. 

\begin{figure}
\centering
\includegraphics[width=0.7\columnwidth]{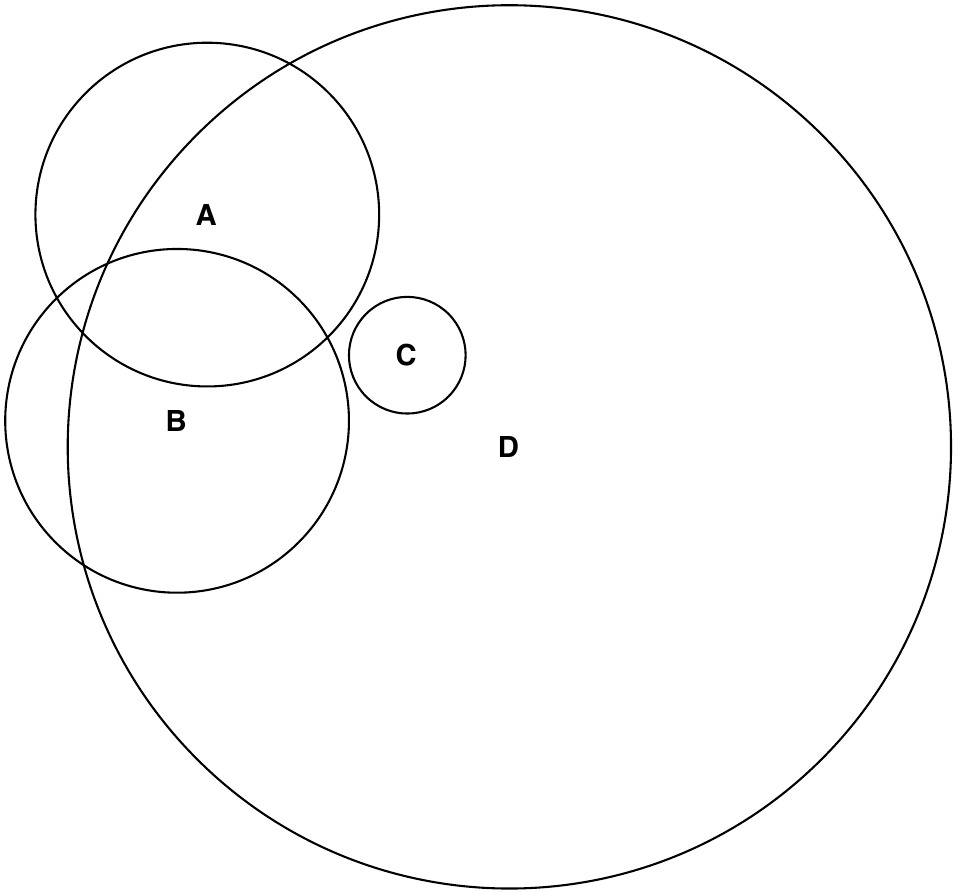}
\caption{Nodes A, B, C and D with coordination areas.}
\label{fig:f_example}
\end{figure}

To illustrate how the discovery mechanism works, Figure~\ref{fig:f_example} depicts four radio devices with omnidirectional antennas, A, B, C and D, seen from above. The circles around the devices are their respective coordination areas, as determined by the propagation model (discussed later). We can see that nodes A and B may cause interference in the same area and should therefore know about each other. C on the other hand, is only within D's coordination area and can safely be ignored by A and B. D is a strong radio transmitter and interferes with all the other nodes. From this example we can see that although D is farther away from A and B, D is much more important in terms of resource allocation than C. The discovery protocol lets A, B and D find each other's IP-addresses, which enables them to coordinate their spectrum use over the Internet.

To set the coordination area of the discovery protocol, the nodes must use a simplified propagation model to estimate its size based on radio parameters. We propose a simple equation to obtain the coordination area radius based on how far a radio signal sent from the node can go before it is inseparable from the general noise in the area. This requires estimations of both noise and signal propagation that are \emph{inaccurate}, but the system requirement is only that \emph{at least} all nodes that we need to coordinate with are included in the set of candidate nodes. More accurate calculations are then later performed by the resource allocator. Conservative values should be used in the estimation, as it is better to discover too many than too few nodes. 

To set the coordination area one can assume that each node has a maximum transmit power $P_i$ and an environmental noise floor $Z(x_i,y_i,z_i) = Z_i$. The noise floor constitutes the sum of all noise sources such as thermal noise and background interference, and can also include location dependent noise sources. The background interference is the interference that a node cannot mitigate through coordination and is thus a result of the sum of interference from far-away nodes. This will depend on network density and location or can be estimated based on stochastic geometry. As there is no need to coordinate with nodes that are sufficiently far away such that their transmit power is below the noise floor, we can find the coordination area radius by using a given propagation model that describes how power is attenuated with distance. Assuming a simple propagation model where power is attenuated according to a power law with path-loss exponent $\alpha$, the equation for the coordination area radius ($cr$) is given as

\begin{equation}
cr_i = \biggl(\frac{P_i}{Z_i}\biggr)^{1/\alpha}.
\end{equation}

\begin{figure}
\centering
\includegraphics[trim=3cm 2cm 3cm 1cm, clip=true, width=0.7\columnwidth]{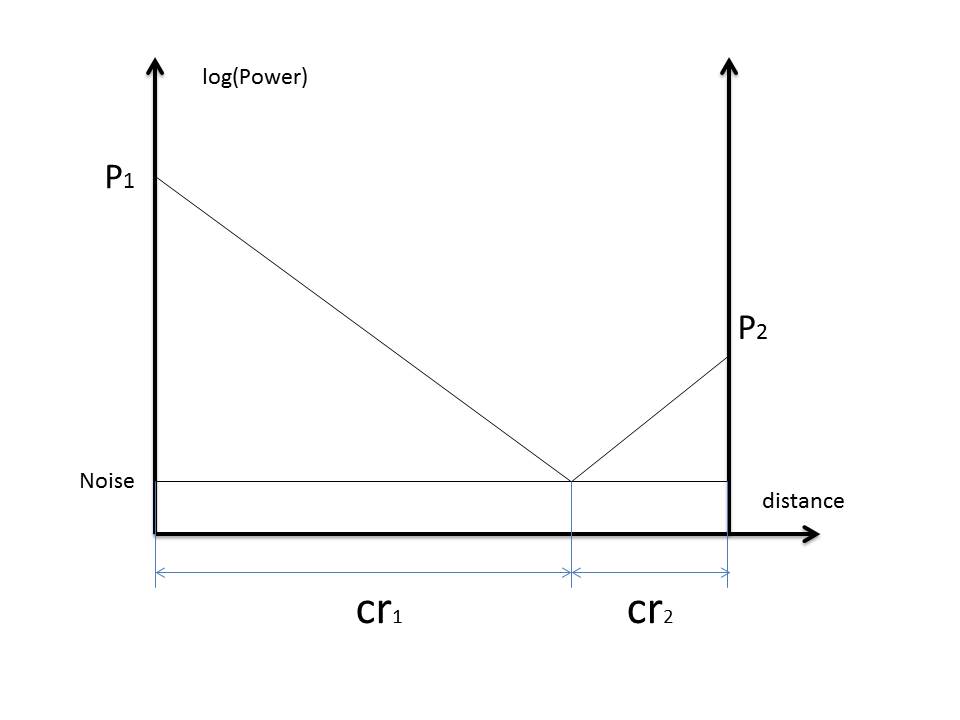}
\caption{Coordination area radius ($cr$) set based on power ($P$) and accumulated noise. Note that signal strength attenuates linearly in this plot as SNR is given in dB, which corresponds to a power-law attenuation given with a path-loss exponent $\alpha$.}
\label{fig:coordination-range}
\end{figure}

Figure~\ref{fig:coordination-range} illustrates this concept with two nodes, transmitting at power $P_1$ and $P_2$ respectively. The coordination area radius of each node is set where the power approaches the noise floor. Here we have assumed that the noise floor is equal for both nodes.

To discover other nodes with overlapping coordination areas, the system also requires that radio nodes know their approximate geographical position. The position is obtained by a location service, such as GPS or Wi-Fi triangulation (e.g. \cite{mozillalocation}), or by letting the user enter a street address or location. Mobile phones with location services could be used to automatically configure wireless devices they are connected to, such as Wi-Fi routers. Advances have also been made in in-door location systems (e.g. ArrayTrack\cite{xiong13}) and we expect that in the near future it will be common for other wireless devices than mobile phones to know their geographical location. An increased coordination area can be used to compensate for inaccurate location information if necessary.

\section{Discovery protocol}
\label{sec:disco}

\begin{figure}[h]
\centering
\includegraphics[width=0.8\columnwidth]{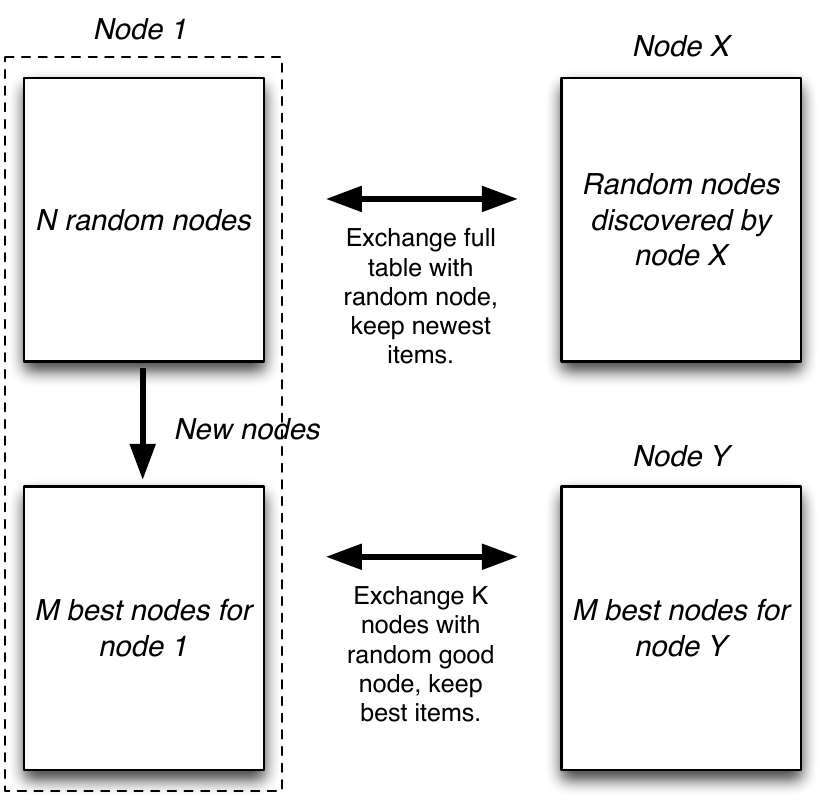}
\caption{Overview of the discovery protocol.}
\label{fig:p2p_tabeller}
\end{figure}

The discovery protocol is based on an unstructured P2P overlay with two mechanisms. The mechanisms are executed periodically, for example four times per minute. The first mechanism maintains a random sample of all nodes participating in the network using a peer sampling protocol~\cite{jelasity07}. When the sample is updated, the new nodes are sent to the second mechanism. 

The second mechanism maintains a list of the nodes with the highest utility so far, i.e., the most important nodes, using an adapted version of the T-Man~\cite{jelasity09} protocol. Utility is calculated based on distance and degree of overlap (defined in Section~\ref{sec:utility}). The most important nodes are periodically contacted to exchange information about other, even more important nodes they may have discovered. This allows the set of important nodes to gradually improve. Finally, the important nodes that have overlapping coordination areas (utility $>= 1$) are added to the set of candidate nodes. The set of candidate nodes is the output of the protocol. An overview of the mechanisms is presented in Figure~\ref{fig:p2p_tabeller}. 

In the following we describe the protocol in more detail.

\subsection{Random sampling}
\label{sec:random}
In order to provide an approximate random sample of the network we use a gossip-based peer sampling service. In our implementation, we have used the Newscast protocol from \cite{jelasity03}, but other peer sampling services could also be used. 

The Newscast protocol maintains a table of $N$ data objects, called \emph{news items}. Each news item represents a node in the network. The news item contains the IP address of the node that produced it as well as a timestamp for when it was created. The timestamp is used to update or remove old items. $N$ is typically much smaller than the number of nodes in the network. In our simulations, we have used $N$ ranging from 5 to 40.

At periodic intervals, a random node is selected and sent the full table of $N$ news items. The sender's own news item is also included with an updated timestamp. The selected node replies with its own table and news item. The table in both nodes now contain $2N$ news items. To reduce the length of the tables, the oldest entries are deleted until the length is again $N$. 

The process is then repeated with a new randomly selected node. By using this mechanism, nodes have a near random set of other nodes participating in the P2P network. 

To enable discovery based on location, we include several fields in each news item in addition to the IP address and the timestamp. These fields are: A randomly assigned source node identifier, the geographical location of the source node, and the radius of the coordination area. The source node identifier is used to enable multiple devices to use the same IP address, while the location and radius are used for discovery. 

Table~\ref{tab:newsitem} contains the complete list of fields distributed with the random sampling mechanism. Note that the lengths suggested in the table are conservative values and could be reduced in many cases. For example, the identifier does not have to be 8 bytes if it is known that only a few instances will be running on each IP address.

\begin{table}
	\centering
	\begin{tabular}{|p{2cm}|p{1cm}|p{3.5cm}|}
	\hline
        \emph{Field} & \emph{Length} & \emph{Description} \\ \hline
		Identifier & 8 bytes & Overlay node ID \\ \hline
		Location & 16 bytes & Geographical location \\ \hline
		Coordination radius & 8 bytes & Radius of coordination area \\ \hline
		IPv4 or & 4 bytes & Source IP \\ 
		IPv6 address & 16 bytes & \\ \hline
		Timestamp & 8 bytes & When the news item was created \\ \hline
	\end{tabular}
	\caption{Fields included in news items. Total length with IPv6 is 56 bytes.}
	\label{tab:newsitem}	
\end{table}

\subsection{Utility function}
\label{sec:utility}
The nodes we are interested in coordinating with are other nodes that have coordination areas that overlap with our own, as illustrated in Figure~\ref{fig:f_example}. This cannot be solved with a regular distance function, as nodes far away may overlap with us, while a node right next to us may have a very small coordination area and not overlap at all. If we have to choose between two non-overlapping nodes however, we would prefer to have contact with the node that is closest to us. This is because a closer node is more likely to have gathered information in the P2P system that is relevant to us.

We define the utility function based on the observation that if the sum of the radii of two coordination areas is higher than the distance between the nodes, their coordination areas overlap. A utility function based on the ratio of the sum of the radii over the distance will thus be $>=1$ when the areas are overlapping. If the areas are non-overlapping, the utility will be $<1$ and approach $0$ as the nodes move farther apart.

This utility function is shown in Equation~\ref{equ:f}, where $x_{i,j}$, $y_{i,j}$, $z_{i,j}$ and $cr_{i,j}$ are the locations and coordination area radii for nodes $i$ and $j$, respectively. When the areas are overlapping $f()$ is higher than 1, while with no overlap the result is less than 1. When the distance between $i$ and $j$ increases, $f()$ approaches $0$.

	\begin{align}
		f(x_i, y_i, z_i;&  x_j, y_j,z_j; cr_i, cr_j) = \nonumber \\
		 &\frac{(cr_i + cr_j)^2}{(x_j - x_i)^2 + (y_j - y_i)^2+(z_j - z_i)^2}
			\label{equ:f}
	\end{align}

Note that the distance used here assumes a flat geographical area, which is practical in a simulated system. In a real system, the distance calculation would have to take earth's shape into account, for example by using the haversine formula, which is the approach we use in our Wi-Fi prototype.

\subsection{Important nodes}
\label{sec:importantnodes}
Important nodes are the nodes discovered so far that have the highest utility. The important nodes are stored in a table that has a fixed length $M$. The nodes in the table are represented by the same news item as is used by the random sampling protocol. When the table is full, nodes with low utility are removed first. As nodes are periodically added from the random sample (described in Section~\ref{sec:random}), the $M$ nodes which have the highest utility are eventually discovered. However, waiting for all candidate nodes to be discovered randomly can take time in large networks. 

To reduce the discovery time, nodes in the table of important nodes are also periodically contacted. This is based on the assumption that a node with high utility is likely to have information about other nodes of interest to us in the same area. 

The node to contact is selected from nodes with overlapping geographical areas, i.e., with utility $>= 1$. If there are fewer than 10 nodes that overlap, we select a node from the top 10 nodes with the highest utility. The node can be selected randomly, but in our simulations the strategy that gave the shortest discovery times is to prioritize selected nodes by 1) utility and 2) longest time since last contact. This strategy ensures that we always attempt to contact newly discovered nodes with high utility first, and then proceed to contact older nodes sorted by longest time since last contact.

After the node is selected, we extract a subset of $K$ nodes from our table that have the highest utility \emph{as calculated for the selected node}. We proceed to send this subset to the selected node. In return, the selected node replies with a list of $K$ nodes it has calculated to have the \emph{highest utility for us}. Finally, both nodes merge the list of $K$ nodes with their table of $M$ entries, deleting the least important entries.

If the number of overlapping nodes exceeds $M$, i.e., a node has more than $M$ nodes with a utility higher than 1, the nodes are not able to discover all their candidate nodes. This is likely to occur if the number of overlapping nodes can not be estimated in advanced, resulting in $M$ being set too low. The table may then be extended dynamically, for example in fixed intervals of 50 nodes to reduce the number of resize operations required before the correct table size is found. This does not affect the bandwidth use, as $K$, which determines the bandwidth used in each periodic interval, is left unchanged.

Another optimization we observe to give shorter discovery times in large networks is to not strictly delete nodes by lowest utility, but to always keep a small set of nodes that are far away (has low utility). This enables other nodes that contact us on their search for overlapping nodes to quickly ``jump'' across large distances in the topology.

We implement a weighted delete mechanism by measuring the distance between the borders of the coordination areas for nodes that have utility $< 1$, i.e., that are not overlapping with us. We then use $log_2$ of the distance to classify the nodes into bins and then balance the number of nodes remembered in each bin. Nodes in the bins with the most nodes are deleted first. As we later see in the evaluation, this optimization decreases convergence time dramatically in some topologies. The idea of maintaining harmonically distributed links to other nodes based on a distance function has previously been used in, e.g., Symphony\cite{symphony} and with T-Man in Vitis\cite{vitis}, although the distance in these works is measured in terms of address space not geographical distance.

To reduce the impact of partitions in the radio topology we also make sure that we attempt to maintain an equal amount of old nodes in four quadrants around each node. Without this optimization, a node at a partition border may end up remembering only members of its own partition, as members of other partitions are too far away to be considered important. By balancing the number of nodes remembered in each of the four quadrants, this effect is avoided. This solution was first suggested (but not implemented) in \cite{jelasity06}.

A final optimization related to the implementation of the protocol is to clear memory in blocks to reduce the number of delete operations. This means that when the table of important nodes is full, we always delete a predefined number of entries that is higher than what we currently need. This reduces the number of times we have to calculate which nodes to delete. The optimization may have a negative effect on the discovery times, as it reduces the number of nodes with utility $< 1$ that is stored in the table, which leads to less information being available to other nodes. The effect is similar to what would happen if we periodically decreased the length of the table ($M$). For example, if we have a table of important nodes with length $M=200$ and a delete block size of 50, the table is effectively reduced to $M=150$ each time the table becomes full. The real table length at a given time is thus somewhere between $M=150$ and $M=200$. As we later show in the evaluation (Section~\ref{sec:eval}), the performance gained by increasing M above 150 is limited and the difference in time between $M=150$ and $M=200$ is thus expected to be minimal for networks with similar densities.

\subsection{Hardware, memory and bandwidth considerations}
\label{sec:bw}
The memory requirements of the protocol are bounded by size $N$ of the random sample and the number of remembered nodes $M$, as well as the size of each news item (see Table~\ref{tab:newsitem}). As we also need buffers for receiving updates from other nodes, the total memory requirement is approximately $2N + M + K$. In the evaluation in Section~\ref{sec:eval}, we see that the protocol reaches a stable state with relatively low values for $N$, $M$ and $K$, 20 and 400 and 200 respectively. If we assume that each news item is close to 56 bytes long, the total memory requirements for storing the data in the tables would be 15680 bytes. This should make the protocol suitable for implementation in modest hardware.

The bandwidth consumption is determined by the length of the periodic interval, the size of $N$ and the size of $K$. If we assume $N=20$ and $K=40$, the data sent back and forth at each interval would be 6720 bytes. With a periodic interval of 15 seconds, the average amount of data sent and received from a single client would be approximately 0.5 kilobytes per second. In terms of Internet traffic where traffic is measured in megabits, this is fairly low. By decreasing $K$ or having a longer periodic system update interval, the average bandwidth consumption can be reduced at the expense of a less responsive system.

\subsection{Resource Allocator Considerations}
\label{sec:ra_introduction}
The discovered nodes that have utility $>=1$ are delivered as a set of candidate nodes to the resource allocator. The resource allocator receives the network address, coordination area and location of each candidate node - everything that is included in the news item, as shown in Table~\ref{tab:newsitem}.

It is now the responsibility of the resource allocator to reduce the candidate node set to only include nodes it has to take into account during resource allocation. The resource allocator may contact resource allocators running on other nodes to gather additional information or perform coordination.

Generic information about the radio devices, such as which wireless access standard a radio node uses, can also be added to the news item in the P2P protocol, and thus be included in the candidate node set. The news items are however slowly updated across the network, and adding parameters to it would affect the bandwidth use of the protocol. The news item should therefore not be used to transfer time critical information or large information items.

Due to the distributed nature of the discovery mechanism, the resource allocator does not know when the set of candidate nodes is complete. The resource allocator should take this into consideration and be able to make more intelligent decisions related to the network resources as more nodes are discovered. An example of a frequency allocation scheme which benefits from the gradual increase in knowledge can be found in \cite{Ellingsater2013b}. This algorithm starts as a selfish algorithm but as the algorithm obtains knowledge about nearby radio devices, it weighs its impact on these devices against a selfishly optimal allocation. Not only does this increase the total performance of the radio networks, but it also decreases the variance of the performance for each radio device.

Another consideration is that each node may belong to a different group of candidate nodes. This can be seen in Figure~\ref{fig:f_example}, where node D has A, B and C as candidate nodes, while C only has D. A resource allocation algorithm running on each of the nodes must therefore be able to coordinate spectrum access with only a \emph{local view} of the area it is in.

Resource allocation is an active research area. The protocol proposed in this paper provides a means to establish a control channel (over the Internet), which is a missing link in many solutions. 

\section{Evaluation}
\label{sec:eval}
To evaluate the protocol we implemented a Bulk Synchronous Parallel (BSP) simulator that enabled us to run deterministic, large-scale simulations. The advantage of using a BSP-based simulator compared to event-based simulators is that we can take full advantage of multi-core systems. On a dual 8-core 2.2 GHz Xeon E5-2660 processor system we are able to simulate networks with more than 1 million nodes with an average of about 60 seconds per iteration.

A limitation of this approach however, is that we are only able to measure time in iterations, not in fractions of seconds, as is common in event-based simulators. Another limitation is that nodes are only allowed to receive messages before each iteration starts and send messages after an iteration ends. This is what enables the simulator to run internal node calculations in parallel within each iteration, as long as messages sent between the nodes are delivered deterministically afterwards. As a consequence, a node needs two iterations to receive a reponse to a message it has sent. To see this, consider the following example. A node sends a message in iteration 1. The message is delivered to its recipient by the simulator in iteration 2. The recipient sends a reply, which is delivered in iteration 3. 

To compensate for this phenomenon, we configured all nodes to alternate between sending requests and responding to requests every second iteration. In other words, two iterations in the simulator corresponds to one request/response-cycle. 

To evaluate the protocol, we first look at how the protocol performs in random, uniform networks. We start by evaluating the effect of adjusting the main parameters $K$ and $M$, before we evaluate some of the optimizations described in Section~\ref{sec:importantnodes}. We proceed to look at increases in bandwidth use and convergence time as the number of nodes in the network increase. Finally, we evaluate the performance under continuous churn in topologies generated from population patterns in the four municipalities Vinje, Tynset, Lillehammer and Oslo and the country of Norway.

\begin{table}
	\centering
    \begin{tabular}{| l | l | l |}
	\hline
    \multirow{6}{*}{}Vary K/M& Nodes & $2^{16}$ \\ 
    (Section~\ref{sec:varykandm})& CA radius & 2-50 m \\ 
           & CN density & 12, 25, 39 \\
           & Protocol (Vary K)& K=10-200, M=200, N=5 \\ 
           & Protocol (Vary M)& K=40, M=100-400, N=0 (dis.) \\ 
           & Figure(s) & Figure \ref{fig:varykit}, \ref{fig:varykbw} and \ref{fig:varymit} \\  \hline
   \multirow{6}{*}{}Distance & Nodes & $2^{18}$ \\ 
  (Section~\ref{sec:distance}) & CA radius & 25 m \\ 
           & Protocol & K=40, M=100, N=5,20,40 \\ 
           & Node distance & 25 m \\ 
           & Purge strategy & Linear, $log_2$ \\
           & Figure(s) & Figure \ref{fig:logdel} and \ref{fig:logdel2} \\  \hline
   \multirow{5}{*}{}Partitioning& Nodes & $2^{18}$ \\
            (Section~\ref{sec:partitioning}) & CA radius & 2-25 m \\ 
           & Protocol & K=40, N=20, M=100 \\ 
           & Partitions & 512 x 512 nodes\\ 
           & Figure(s) & None \\  \hline
   \multirow{4}{*}{}Scalability &  Nodes & $2^{18}$, $2^{20}$ \\
           (Section~\ref{sec:scalability}) & CN density & 2, 7, 12, 17, 25, 44 \\
           & Protocol & K=40, M=100, N=0,5,20 \\ 
           & Figure(s) & Figure~\ref{fig:densityit} and \ref{fig:densitybw} \\  \hline
   \multirow{5}{*}{}Real world& Nodes & See table \ref{tab:muni} \\
            (Section~\ref{sec:realworld}) & CA radius & 2-25 m \\ 
           & Protocol & K=40, M=100, N=20 \\
           & Churn & 1, 2, 5\% \\
           & Figure(s) & Figure~\ref{fig:churnit} and \ref{fig:churndensity} \\  \hline
	\end{tabular}
	\caption{Simulation parameters.}
    \label{tab:sim}
\end{table}

All simulation parameters are summarized in Table \ref{tab:sim}.

\subsection{Varying K and M}
\label{sec:varykandm}
The protocol's performance is mainly dependent on its three parameters; $K$, $M$ and $N$. Recall that $K$ is the number of important nodes sent in each message to other nodes, $N$ is the size of the random sample exchanged between nodes and $M$ is the size of the table containing the most important nodes found so far. Changing either of these parameters changes the performance characteristics of the protocol.

The objective in this experiment is to examine the implications of varying $M$ and $K$. This is measured by looking at the resources required to converge, i.e., stabilize the P2P network after a topology change. This gives an indication of how $K$ and $M$ affects the relative performance of the protocol, but does not represent real world behaviour. Churn, i.e., how the protocol performs in a constantly changing topology, as well as the effect of varying $N$ are evaluated later in this section.

The experiment is performed by first creating a random topology with $2^{16}$ (65536) nodes, where each node is supplied with a random sample of 5 nodes and an empty set of important nodes. The nodes are uniformly distributed in a geographical area with a coordination area radius set randomly between 2 to 50 meters. A simulation is then run until all nodes have successfully located all their candidate nodes. We proceed to repeatedly insert a single node in the network and measure the time and bandwidth required before both a) the node has discovered all its candidate nodes and b) other nodes have discovered the new node.  New nodes are inserted within 50 meters of an existing node geographically, so that they are likely to have at least one candidate node.

The random topologies are generated with different average candidate node densities by varying the size of the geographical area, approximately 12, 25 and 39. The density determines how many nodes each node must discover on average. The simulation is repeated on 10 random topologies for each candidate node density, and the results from inserting 100 single nodes are measured in each topology.

The random samples add new information to the table of important nodes in each iteration. When $M$ is much larger than the number of nodes with utility $>=1$ it will thus gradually contain lot of random data during the simulation. This increases the chance of being able to reply to other nodes with information with high utility. This is however not caused by the gradually improving discovery mechanism, but is an artifact of high memory use in a small topology. As we want to only observe the effect of varying memory use, the random sampling is thus disabled in the simulations where we vary $M$. 

\begin{figure}
\centering
\includegraphics[trim=1cm 1cm 1cm 1cm, clip=true, width = 0.8\columnwidth]{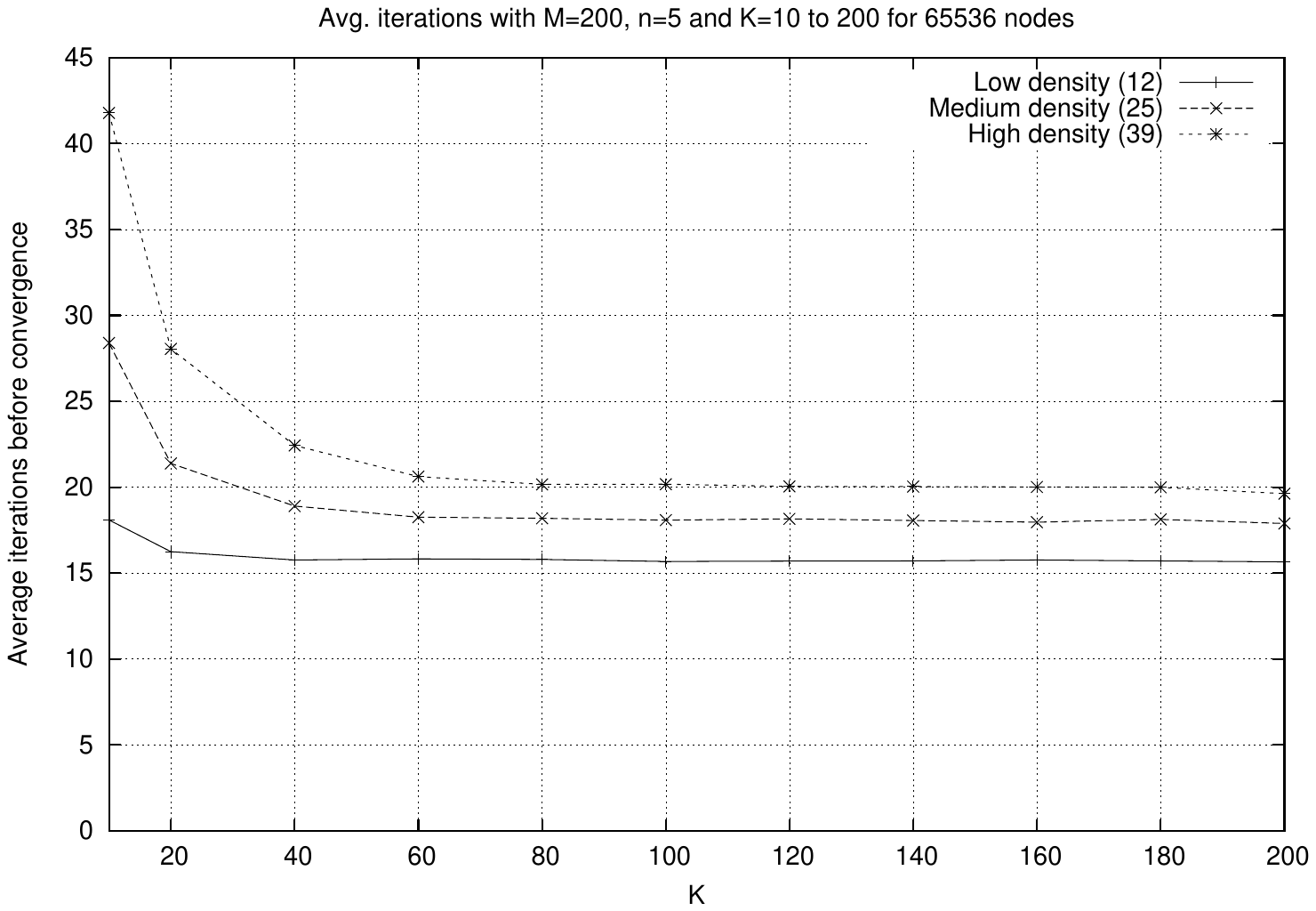}
\caption{Average convergence time when varying $K$}
\label{fig:varykit}
\end{figure}

\begin{figure}
\centering
\includegraphics[trim=1cm 1cm 1cm 1cm, clip=true, width = 0.8\columnwidth]{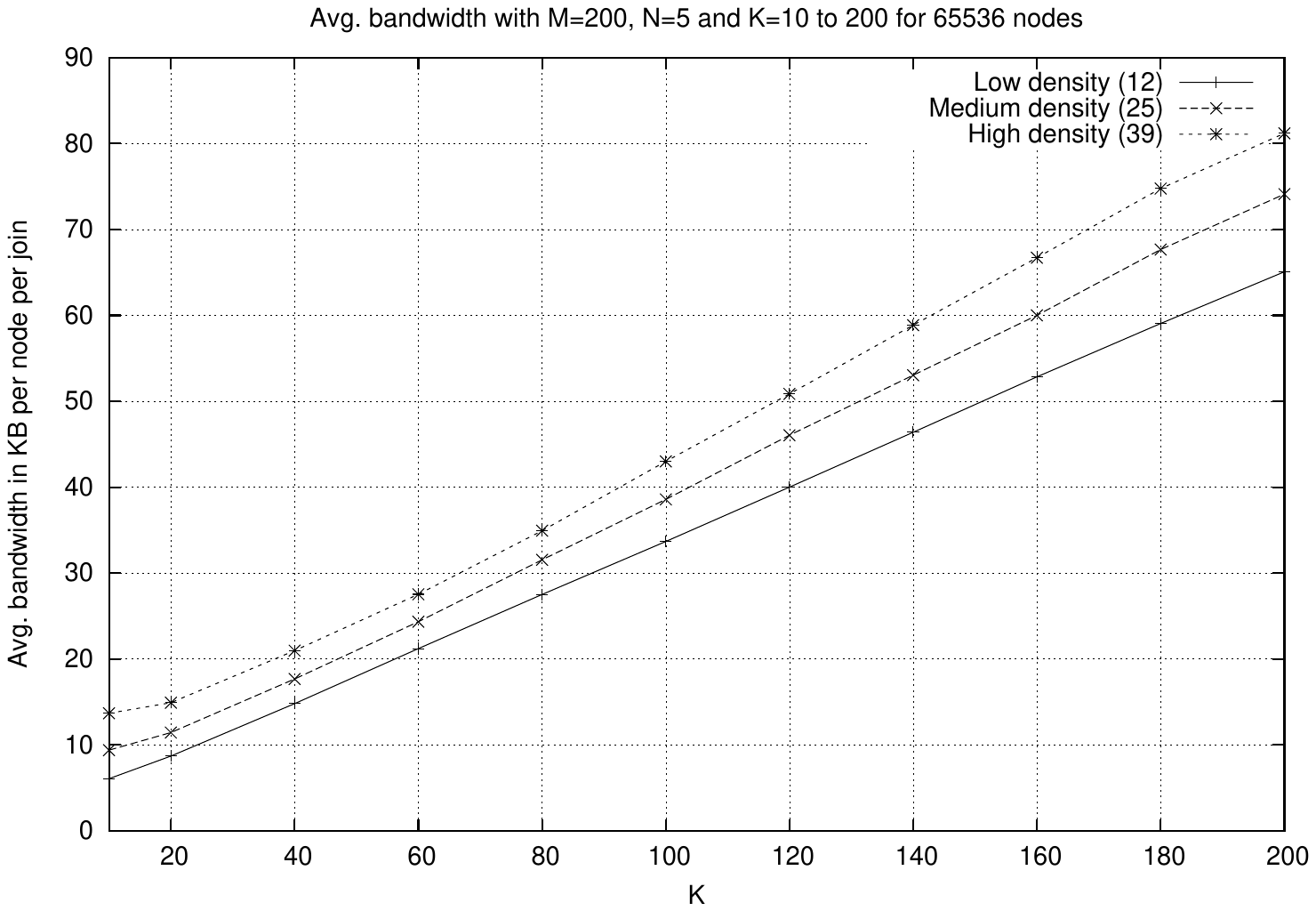}
\caption{Average bandwidth use for varying $K$}
\label{fig:varykbw}
\end{figure}

The average number of iterations for the network to stabilize after a new node has joined is shown for varying $K$ in Figure~\ref{fig:varykit}. As we can see, the number of iterations before convergence decreases as $K$ increases. This is not unexpected, as the larger $K$ is, the more information can be exchanged in each iteration. 

In Figure~\ref{fig:varykbw} the average bandwidth use node in the same experiment is displayed. Interestingly, even if low values for $K$ results in more iterations before convergence, the messages that are exchanged are smaller, which results in lower bandwidth use. There is thus a trade-off between convergence time and bandwidth, with bandwidth use increasing almost linearly with $K$ in our simulations.

\begin{figure}
\centering
\includegraphics[trim=1cm 1cm 1cm 1cm, clip=true, width = 0.8\columnwidth]{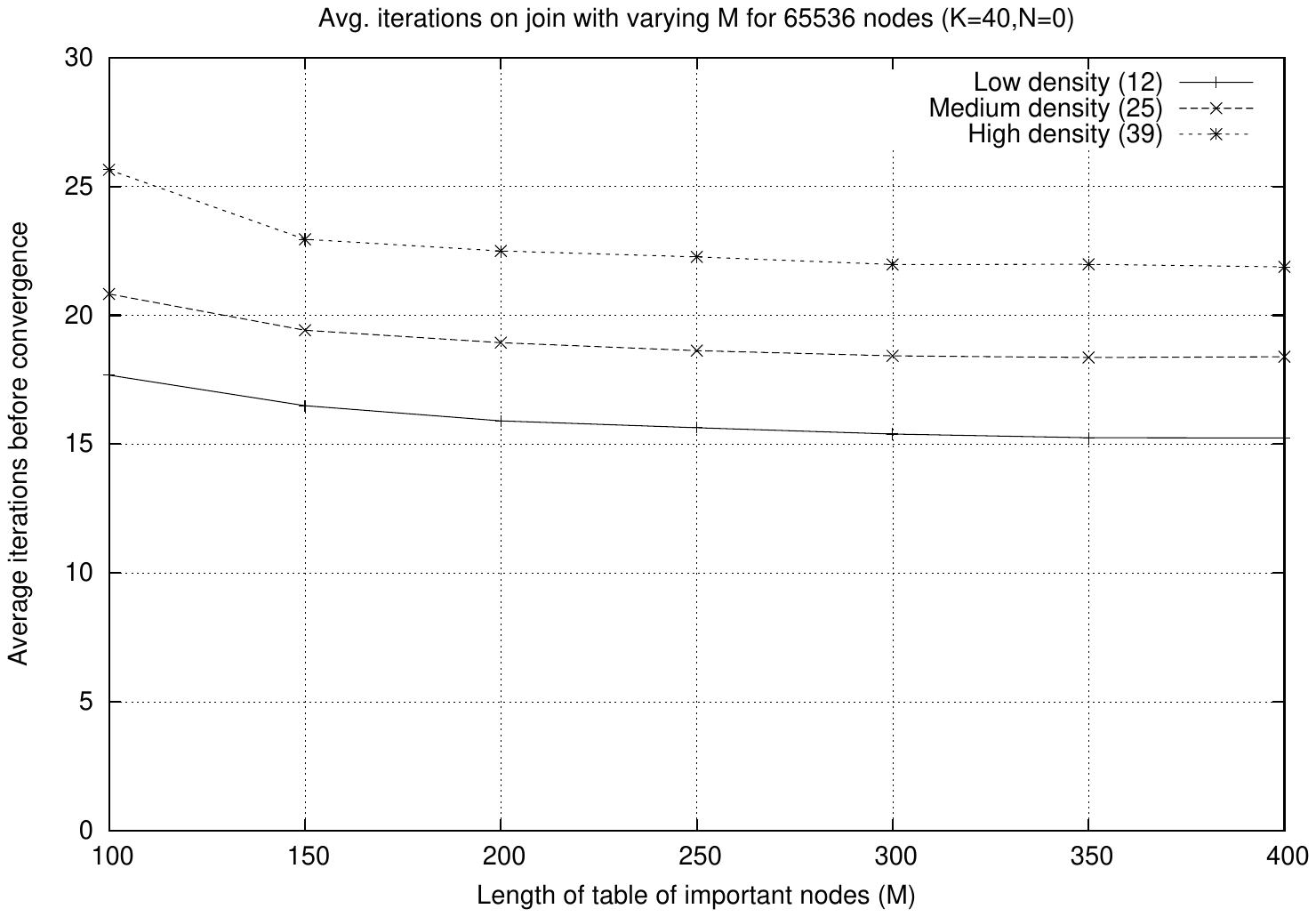}
\caption{Average convergence time when varying $M$}
\label{fig:varymit}
\end{figure}

We repeated the experiment with a fixed $K=40$ and varying $M$ from 100 to 400 to see how memory use affects convergence time. The result is shown in Figure~\ref{fig:varymit}. As when we varied $K$, the improvement in performance subsides after the table reaches a certain length. We can see also see that for the evaluated densities, the improvement in convergence time with $M > 150$ is limited.

\subsection{Distance}
\label{sec:distance}
As the protocol is based on gradually being able to find nodes that are closer to the area one is looking for, the distance between the node one initially connects to and the candidate nodes one is looking for affects the discovery time. This may especially have an impact in a global P2P network, where one may initially be connected to a node far away, e.g., on a different continent.

The purpose of this simulation is to see how node distance and the size of the random sample ($N$) affects convergence time. We also examine how the logarithmic delete optimization described previously (see Section~\ref{sec:importantnodes}) can be used to reduce the need for a high $N$, effectively lowering the bandwidth requirements of the protocol.

For this simulation we created a topology consisting of $2^{18}$ (262 144) nodes on a straight line, where each node has two candidate nodes; the node preceding it and the node succeeding it on the line. The nodes are positioned 25 meters apart, resulting in a 6250 kilometers long line of radio nodes. 

When a node joins, it is given a position along the line and is connected to a randomly chosen existing node (in a different position). It is then forced to gradually connect to nodes closer and closer to itself along the line until it finds its candidate nodes. As in the first simulation, we measure the time it takes for all nodes to find their candidate nodes after a new node has been inserted. This topology enables us to compare discovery time to the distance between the node itself and the node it initially connected to, i.e., the distance ``travelled''.

As $N$ determines the size of the random sample a node receives when joining the network it also affects the discovery time relative to the total number of nodes in the topology. To see this, consider that the probability of an undiscovered candidate node being included in a truly random sample of size $N$ is $\frac{N}{X}$, where $X$ is the number of nodes in the topology. As more samples are received, the probability of discovering candidate nodes (or nodes that are close) increases further. As the topology size $X$ increases however, $N$ must be increased accordingly to maintain the same discovery probability. This increases the overall bandwidth requirements of the protocol and does not scale well. 

We performed the evaluation with $M=100$, $K=40$ and $N=5$ to $40$. The experiments are repeated 19 times for each 100 new nodes, resulting in 1900 samples per data point. 

\begin{figure}
\centering
\includegraphics[trim=1cm 1cm 1cm 1cm, clip=true, width = 0.8\columnwidth]{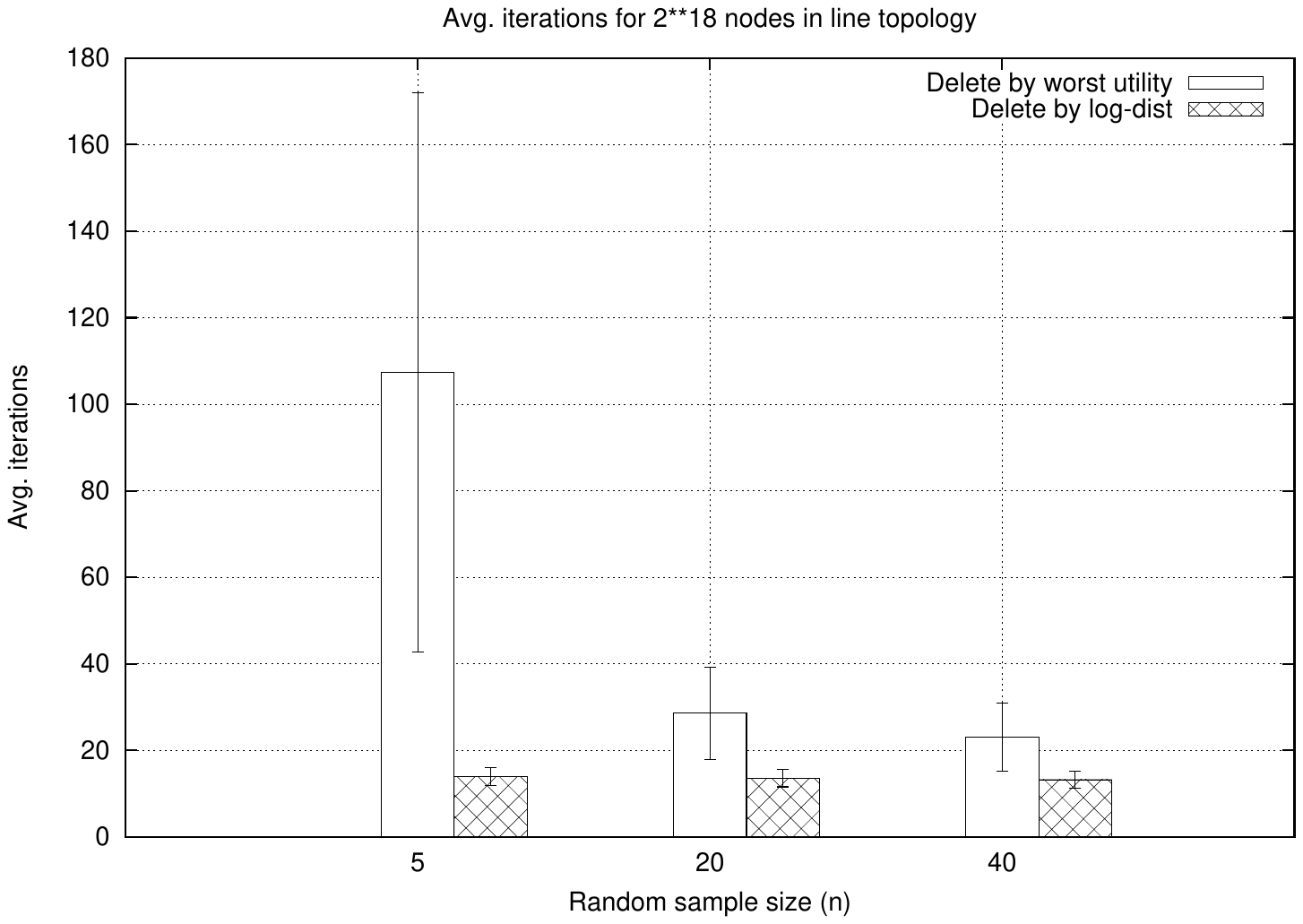}
\caption{Average convergence time when varying N.}
\label{fig:logdel}
\end{figure}

Figure~\ref{fig:logdel} shows the average number of iterations required for convergence after a new node is inserted. We when nodes are deleted from the table of important nodes based on lowest utility, the convergence time largely depends on $N$. As $N$ increases, the convergence time decreases. With the optimization however, the convergence time is consistently low, independently of the size of the random sample. From this one could conclude that the random sample is unnecessary, but it is important to remember that it also functions as an insurance for convergence.

\begin{figure}
\centering
\includegraphics[trim=1cm 1cm 1cm 1cm, clip=true, width = 0.8\columnwidth]{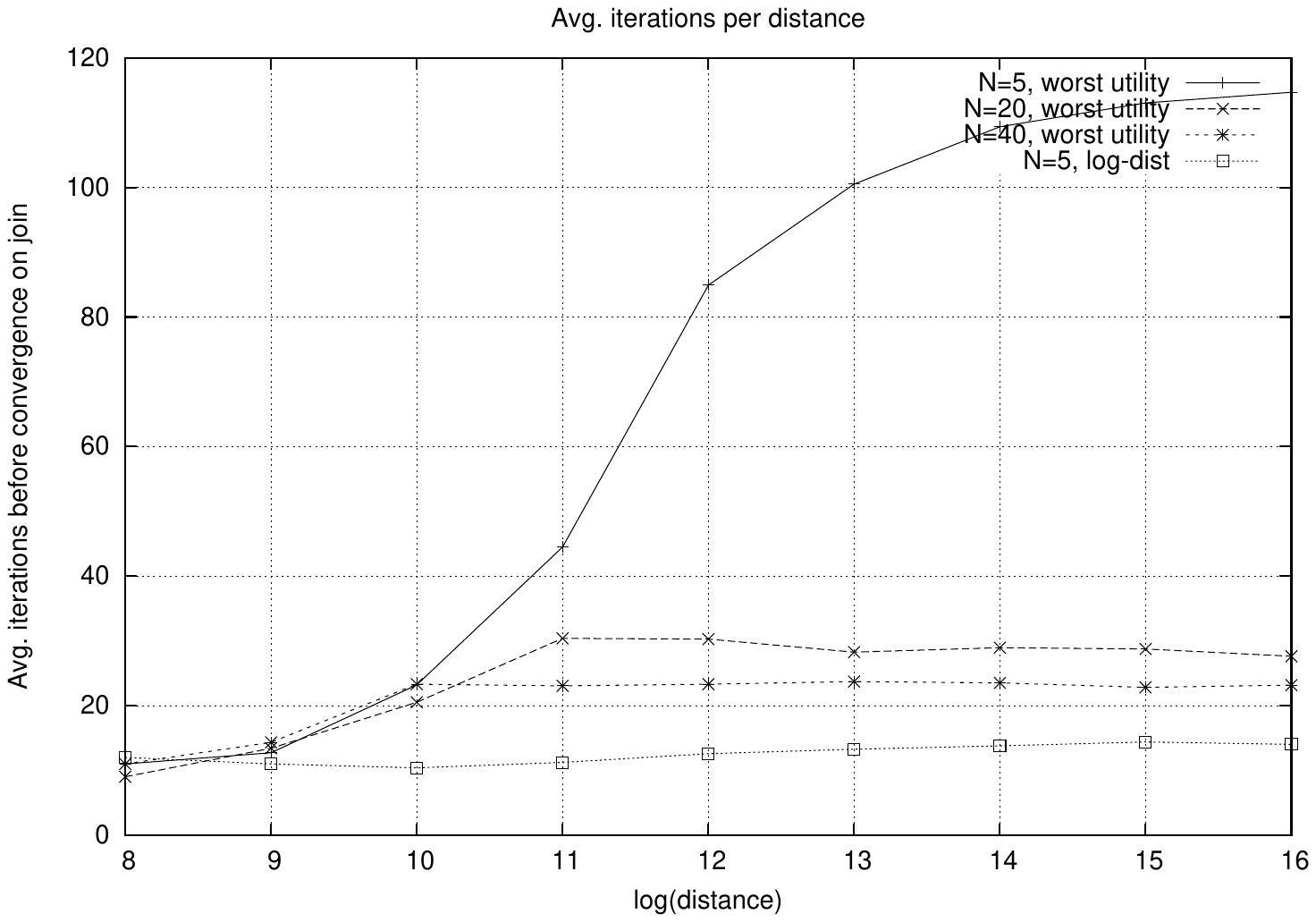}
\caption{Average convergence time over join distance.}
\label{fig:logdel2}
\end{figure}

To further investigate the performance of the optimization, we show the average iterations over logarithmic distance in Figure~\ref{fig:logdel2}. We can see that when $N=5$, the convergence time increases with distance. As N increases relative to the number of nodes in the topology, the convergence time decreases and is more evenly distributed over distance. With the optimization enabled, we get the best results both in terms of bandwidth ($N=5$) and convergence time.

\subsection{Partitioning}
\label{sec:partitioning}
To investigate how robust the quadrant delete optimization (see Section~\ref{sec:importantnodes}) is to partitioning we create a topology consisting of $2^{18}$ nodes separated in 512 groups of 512 nodes. Within the groups, all nodes have overlapping coordination areas, but they never overlap with nodes outside their group. We then compare the time it takes to join with the optimization and without. We disable the logarithmic delete optimization to make sure that each node only knows about other nodes in their partition when the quadrant delete optimization is turned off. We let 100 nodes join separately in 20 random topologies, resulting in 2000 measurements per data point. The experiments were run with $K=40$, $N=20$ and $M=100$. Our results show that without the optimization, the convergence time after join was on average 74.36 iterations with a standard deviation of 69.35. In contrast, the optimization allowed new nodes to join in 17.67 iterations on average with a standard deviation of 6.78. The reason for the long convergence time without the optimization enabled is that when a node joins it is connected to a node that only know about other nodes in its own partition. Nodes on other islands must thus be discovered through the random sampling mechanism, which takes time.

\subsection{Scalability}
\label{sec:scalability}
So far we have only evaluated the protocol with topologies with up to $2^{18}$, or 262 144 nodes. We now evaluate bandwidth use and average join time in topologies with up to $2^{20}$, or 1 048 576 nodes. The goal is to investigate whether the protocol scales by looking at the increase in convergence time and bandwidth use as we add more nodes. 

The experiment is performed as when we varied $K$ and $M$, but here we vary the density of the network and $N$. For $2^{20}$ nodes the experiment is performed with $N=20$ and with $2^{18}$ nodes we have varied $N$ from 0 to 20. Each data point represents 10 randomly generated topologies with 100 separately joined nodes, in total 1000 samples. 

\begin{figure}
\centering
\includegraphics[trim=1cm 2cm 1cm 1cm, clip=true, width=0.8\columnwidth]{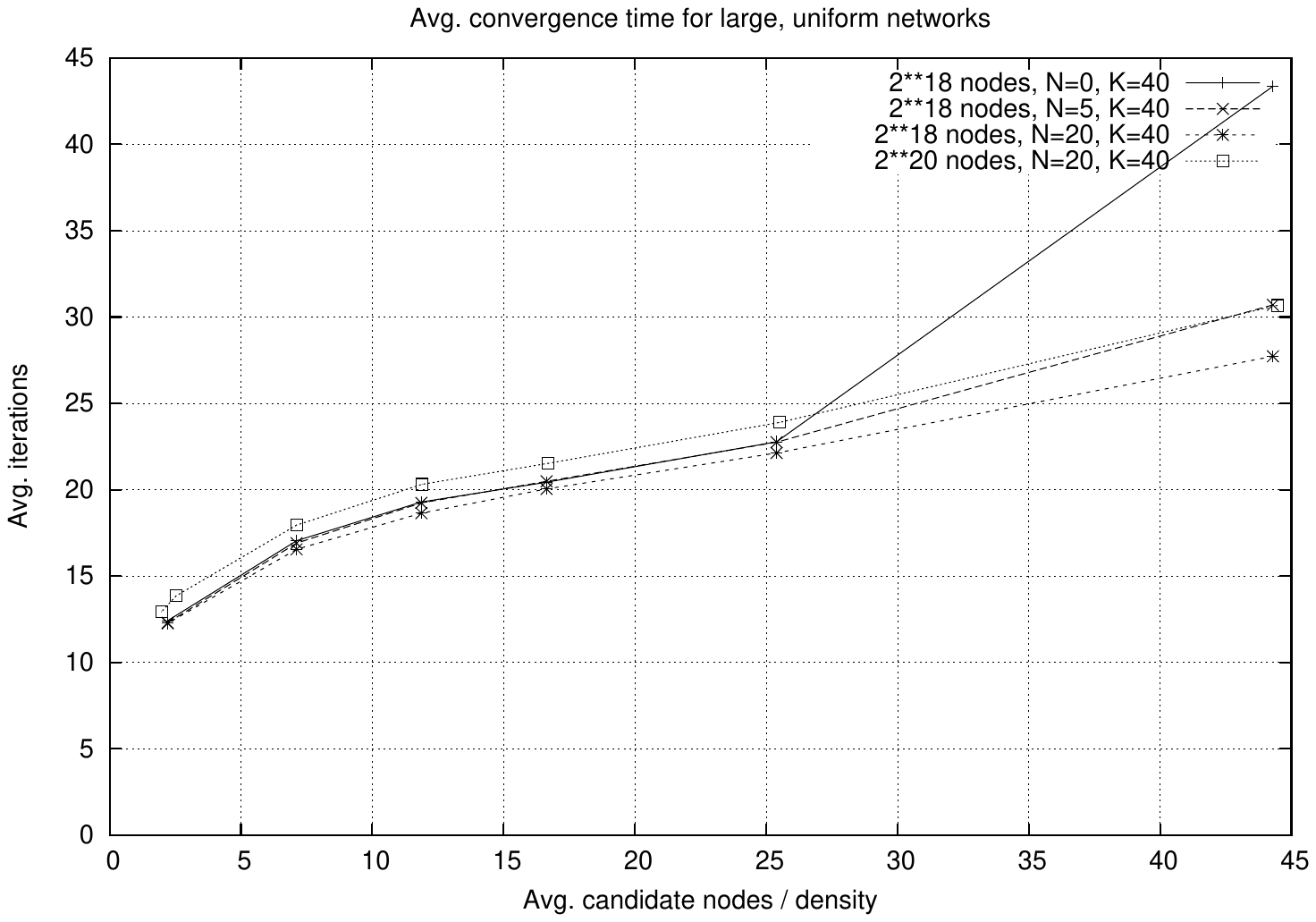}
\caption{Average convergence time over density.}
\label{fig:densityit}
\end{figure}

The average number of iterations required before convergence after adding a new node is shown in Figure~\ref{fig:densityit}. As expected, we can see that as the density increases, the number of iterations required to discover all candidate nodes after a join increases as well. We can also see that the number of iterations required to add a new node only slightly increases from $2^{18}$ to $2^{20}$ nodes. An interesting effect however, is that when the network density increases more than $K$ when $N=0$, the convergence time increases rapidly. This is because the total number of candidate nodes exceeds what can be transferred in a single message (K). When this situation arises, the nodes are configured to select a random subset of $K$ size to fill the message, resulting in some candidate nodes not being included. This increases the convergence time. When $N$ is more than $0$ this effect becomes less dominant as nodes are also randomly discovered. To reduce the discovery time in dense topologies the message size could be dynamically increased to always have room for all candidate nodes, but we have not evaluated this solution here. 

\begin{figure}
\centering
\includegraphics[trim=1cm 2cm 1cm 3cm, clip=true, width=0.8\columnwidth]{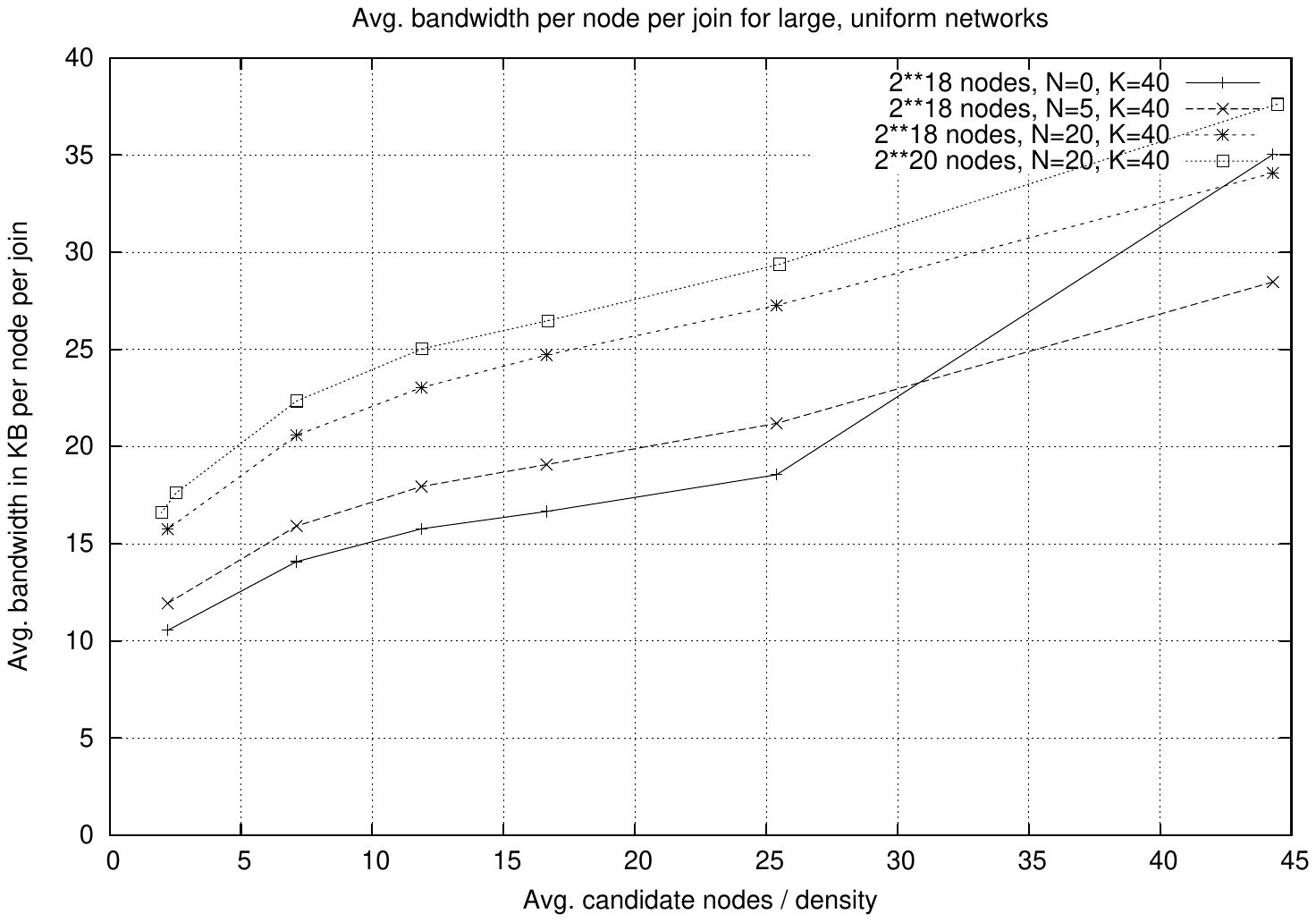}
\caption{Average bandwidth use for join over varying density.}
\label{fig:densitybw}
\end{figure}

In Figure~\ref{fig:densitybw}, the bandwidth results for the same experiment are shown. Again, we can see that even if the number of iterations are high with a low N, the bandwidth use is still relatively low. For the same configuration ($N=40$), $2^{18}$ and $2^{20}$ nodes have a relatively similar bandwidth consumption per node. The slight increase is explained by added iterations required for convergence. Per iteration the bandwidth use is approximately the same.

In our simulations, the protocol thus scales sub-linearly in terms of convergence time and bandwidth relative to the number of nodes in the topology.

\subsection{Real world topologies and churn}
\label{sec:realworld}
In the evaluations so far, the topologies we have evaluated have been uniformly distributed and we have only considered the effect of single topology changes while varying protocol parameters. In a real world application however, it cannot be assumed that the protocol will operate in a stationary environment. Devices will be turned on and off, either due to errors or part of their regular usage patterns. Nodes are also likely to not be uniformly distributed, but clustered around areas where people live.

In this Section, we use data from statistical data for population patterns in Norway~\cite{ssb} to evaluate how the protocol performs in a dynamic environment. The experiment intends to evaluate the performance of the protocol as if it was running on a single device in every home, e.g., in the firmware of a Wi-Fi-router or access point or in a femtocell.

\begin{table}
	\centering
    \begin{tabular}{| l | r | r | r |}
	\hline
    \emph{Topology} & \emph{Nodes} & \emph{Avg. CN} & \emph{Max. CN} \\ \hline
        Vinje & $\approx$ 2 000 & 1.65 & 9 \\
        Tynset & $\approx$ 3 000 & 2 & 12 \\ 
        Lillehammer & $\approx$ 13 000 & 3.6 & 29 \\ 
        Oslo & $\approx$ 312 000 & 15.4 & 117 \\ 
        Norway & $\approx$ 2 338 000 & 3.9 & 90 \\ \hline
	\end{tabular}
    \caption{Overview of topologies generated from population patterns.}
	\label{tab:muni}	
\end{table}

For the simulation, we simulate the performance in a topology with 2.3 million nodes based on the population pattern of Norway. We also simulate smaller topologies based on four municipalities with different population density and size, Vinje, Tynset, Lillehammer, as well as Norway's capital Oslo. The data set for Norway is publicly available for download, while the other four are available on request for research purposes from \cite{ssb}.

For Vinje, Tynset and Lillehammer we have data sets listing the number of inhabitants per 100x100 meter. For Norway we have inhabitants per 250x250 meter. To get the number of homes within the area we divide the number of inhabitants by $2.22$, which is the average number of people per household in Norway. For Oslo we have the actual number of households within cells of 250x250 meters. 

\begin{figure}
    \centering
    \begin{subfigure}{0.48\textwidth}
        \centering
        \includegraphics[trim=2cm 1cm 2cm 1cm, clip=true, width=0.9\linewidth]{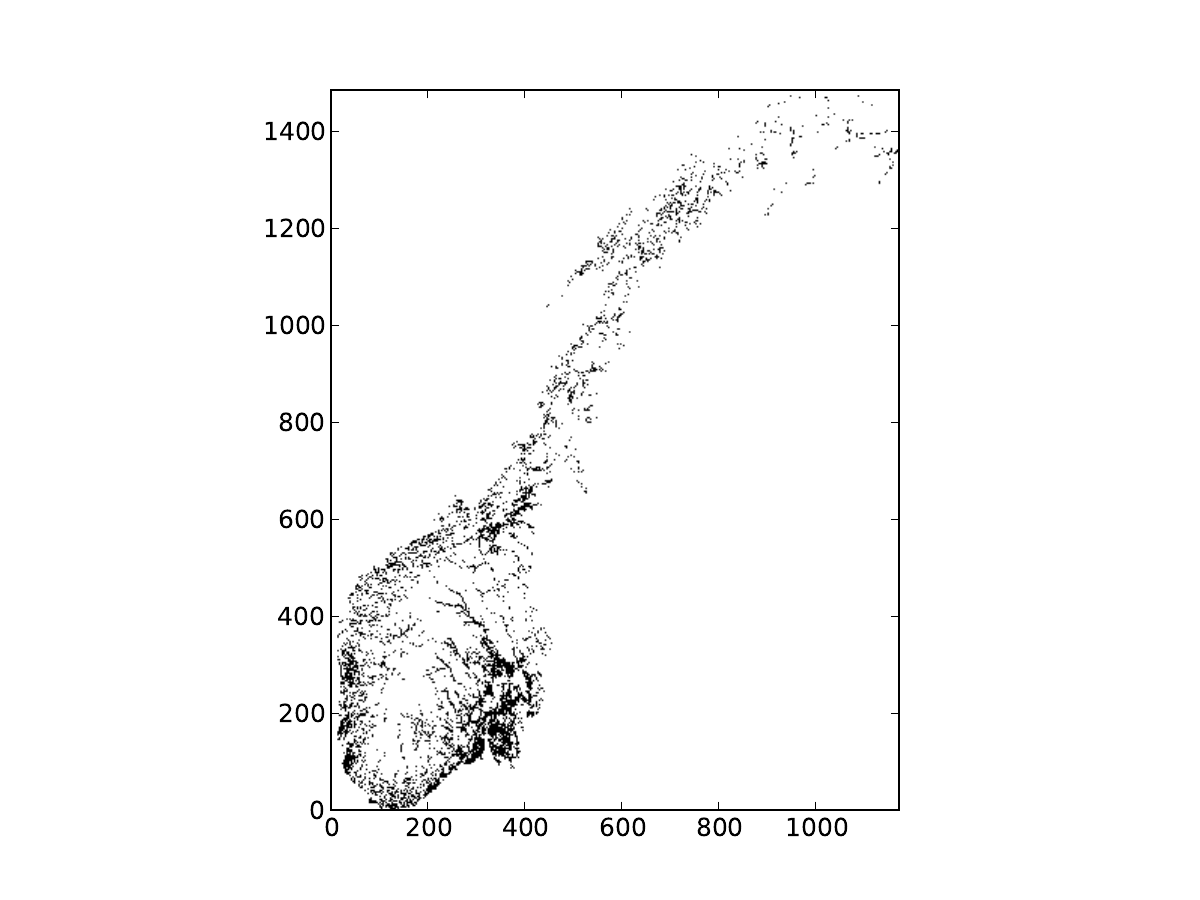}
        \caption{Norway}
        \label{fig:norway}
    \end{subfigure}
    \\
    \begin{subfigure}{0.24\textwidth}
        \centering
        \includegraphics[trim=1cm 1cm 1cm 1cm, clip=true, width=0.9\linewidth]{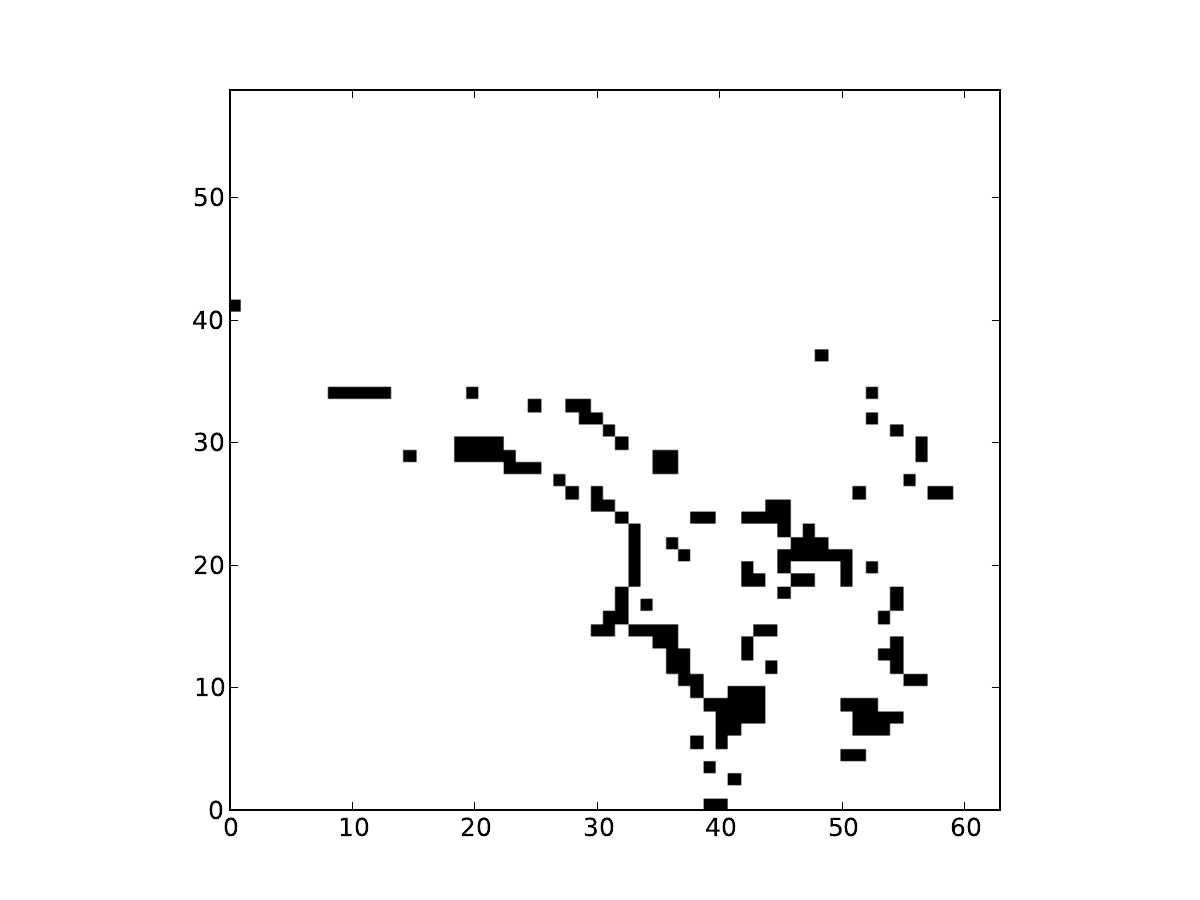}
        \caption{Vinje}
        \label{fig:vinje}
    \end{subfigure}%
    \begin{subfigure}{0.24\textwidth}
        \centering
        \includegraphics[trim=1cm 1cm 1cm 1cm, clip=true, width=0.9\linewidth]{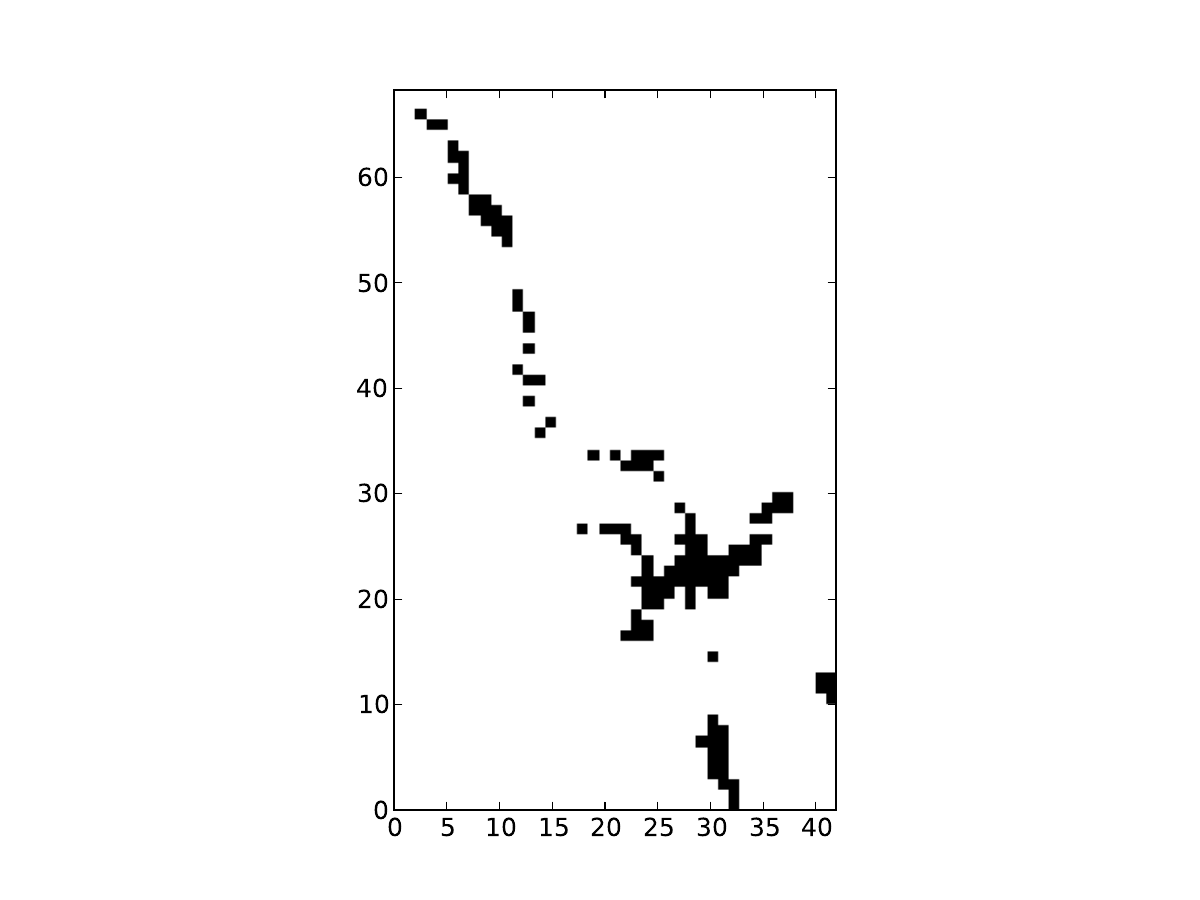}
        \caption{Tynset}
        \label{fig:tynset}
    \end{subfigure}
    \\
    \begin{subfigure}{0.24\textwidth}
        \centering
        \includegraphics[trim=2cm 1cm 2cm 1cm, clip=true, width=0.9\linewidth]{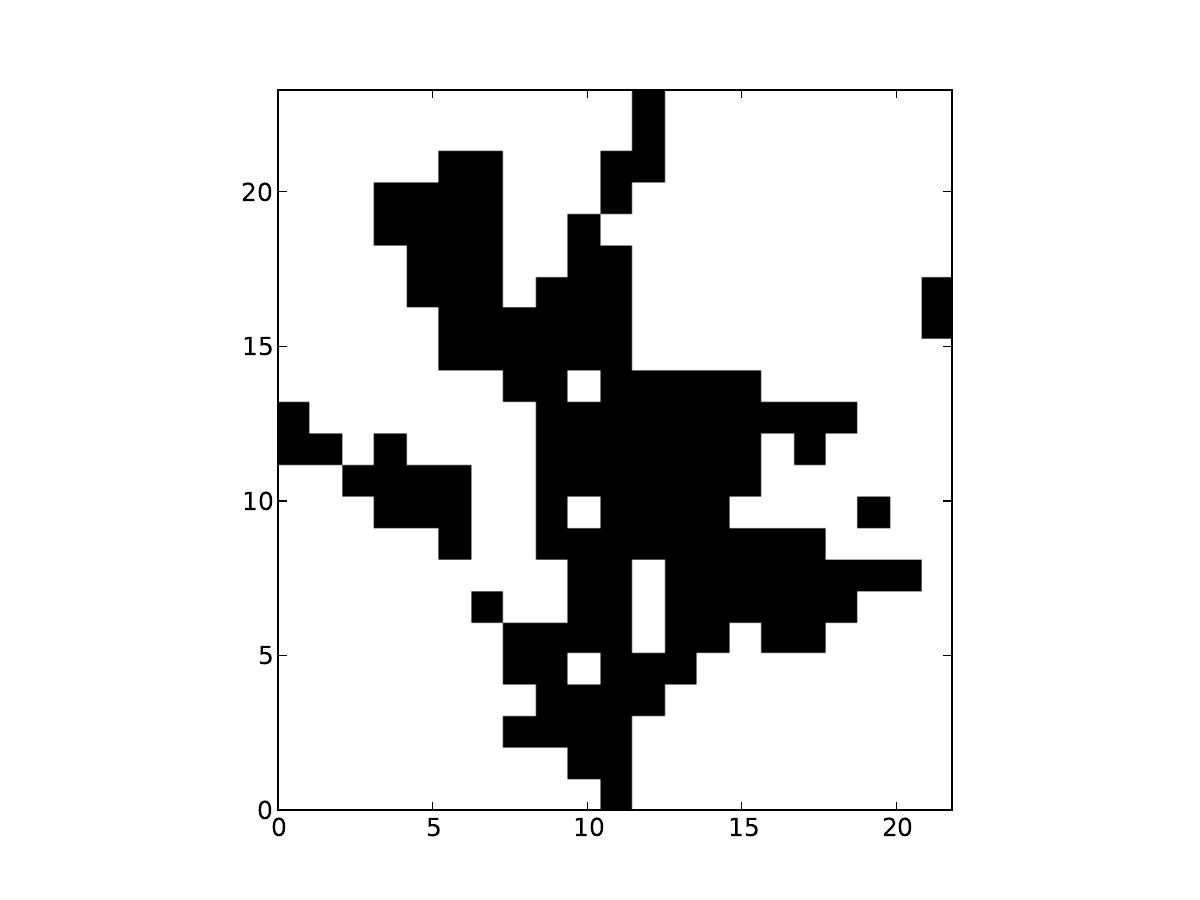}
        \caption{Lillehammer}
        \label{fig:lillehammer}
    \end{subfigure}%
    \begin{subfigure}{0.24\textwidth}
        \centering
        \includegraphics[trim=2cm 1cm 2cm 1cm, clip=true, width=0.9\linewidth]{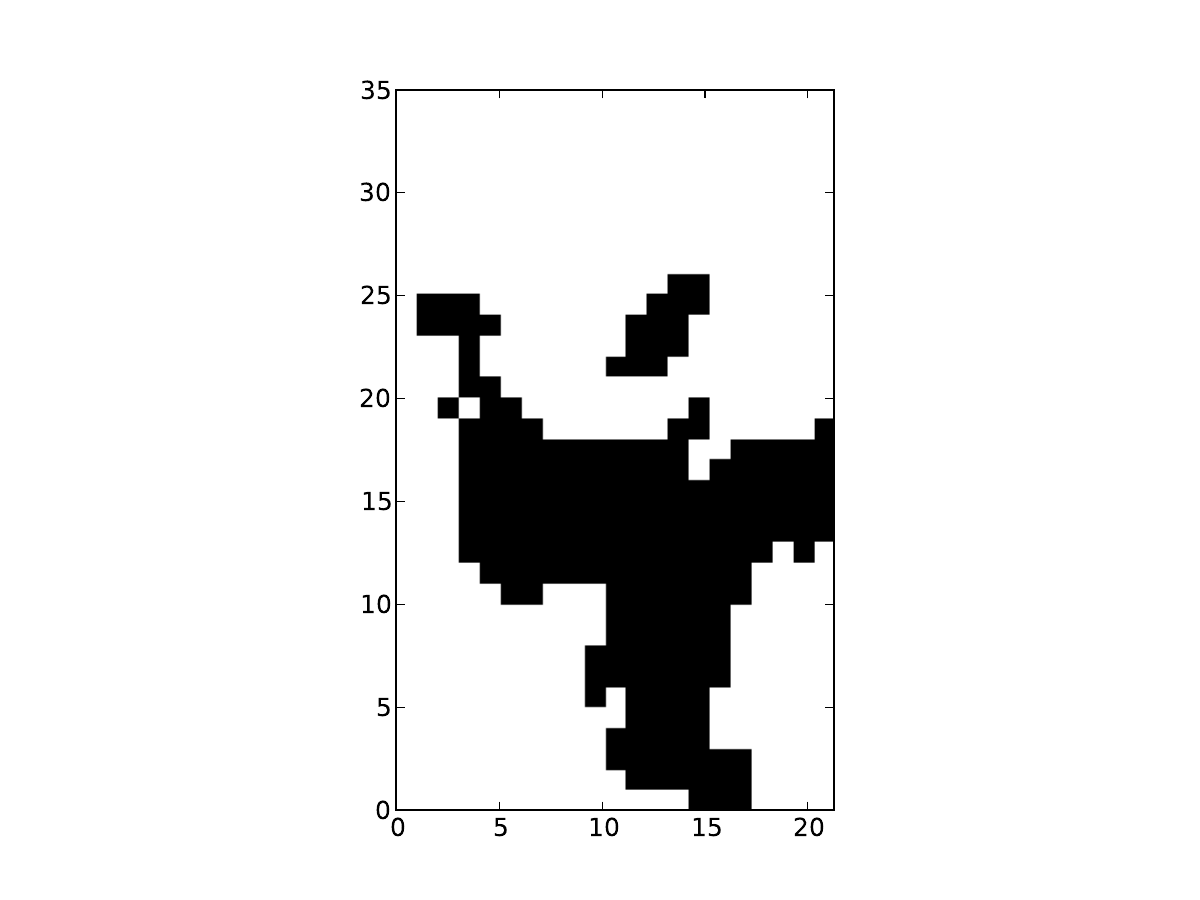}
        \caption{Oslo}
        \label{fig:oslo}
    \end{subfigure}
    \caption{Population patterns}\label{fig:poppat}
\end{figure}

We generated random topologies where each household has a device with a coordination area with a radius between 2 and 25 meters. The number of nodes, approximate average candidate node density and average maximum number of observed candidate nodes are summarized for the topologies in Table~\ref{tab:muni}. Note that as the topologies are generated randomly, the actual values vary slightly from the numbers presented here. An interesting feature of these topologies, and especially Oslo and Lillehammer, is that the distribution of candidate nodes is very long tailed, due to high population density in central areas. The population patterns are displayed graphically in Figure~\ref{fig:poppat} with a resolution of 1 x 1 kilometer. 

Norway has a lower maximum candidate node degree than Oslo, even though Oslo is included in the topology. It is likely that this is caused by the households in Oslo having a lower average number of inhabitants than the country average of $2.22$. As the Oslo data set reflects the actual number of households, this is thus the most accurate topology in terms of household density.

In this experiment, we adapted the protocol to support nodes that leave the network. This was done by adding a timeout of 50 iterations to each entry in the list of important nodes. Recall that each entry has a timestamp attached which is set by the node that originally produced the information item. If this timestamp has not been updated for 50 iterations, it is likely that the node is no longer available in the network and the information can be removed from the list of important nodes. The random sample mechanism is left as it is, as old nodes are automatically forgotten if they fail to send an updated news item. It should be noted that even if this mechanism is based on a global, synchronized timestamp, it is likely that devices are able to synchronize their time with time server on the Internet. They do not need to have the exact time, as long as the clock does not drift more than the timeout value. If time is not available or is difficult to implement on the device, an alternative could be to use \emph{holding time} instead of timestamps, so that each node that holds an item increases a timestamp every second before it passes the item to a new node. This effectively adds an age-counter to each entry, enabling other nodes to determine which items are the oldest and whether they should be removed.

In this experiment we do not evaluate convergence time, as the network never converges. Instead we look at the average ratio of discovered candidate nodes in the network. This represents how accurate a nodes view of the network is when a certain percentage of nodes leave and join the network in each iteration. We evaluate the protocol with 1\% to 5\% churn per minute.

To simulate churn over time, we assume that each node periodically contacts other nodes four times per minute, as in the bandwidth example in Section~\ref{sec:bw}. A minute is then equal to 8 iterations in the simulator, as a full request/response cycle takes two iterations. Within every simulated minute, we select a given percentage of the nodes randomly and disable them. To maintain the size of the network, we add the same number of new nodes. The new nodes are given generated random samples of $N$ existing nodes when they are connected. Nodes are added according to the same geographical distribution as the data set the topology was generated from.

We let the simulation run for 500 iterations for each topology and measure the ratio of discovered neighbors the last half of the simulation (250 iterations). The simulation is repeated with 10 randomly generated topologies per data set, except for Norway which is only run three times due to the size of the topology. The average of the average discovery ratio and the average standard deviation is presented.

\begin{figure}
\centering
\includegraphics[trim=2cm 2cm 1cm 2cm, clip=true, width = 0.9\columnwidth]{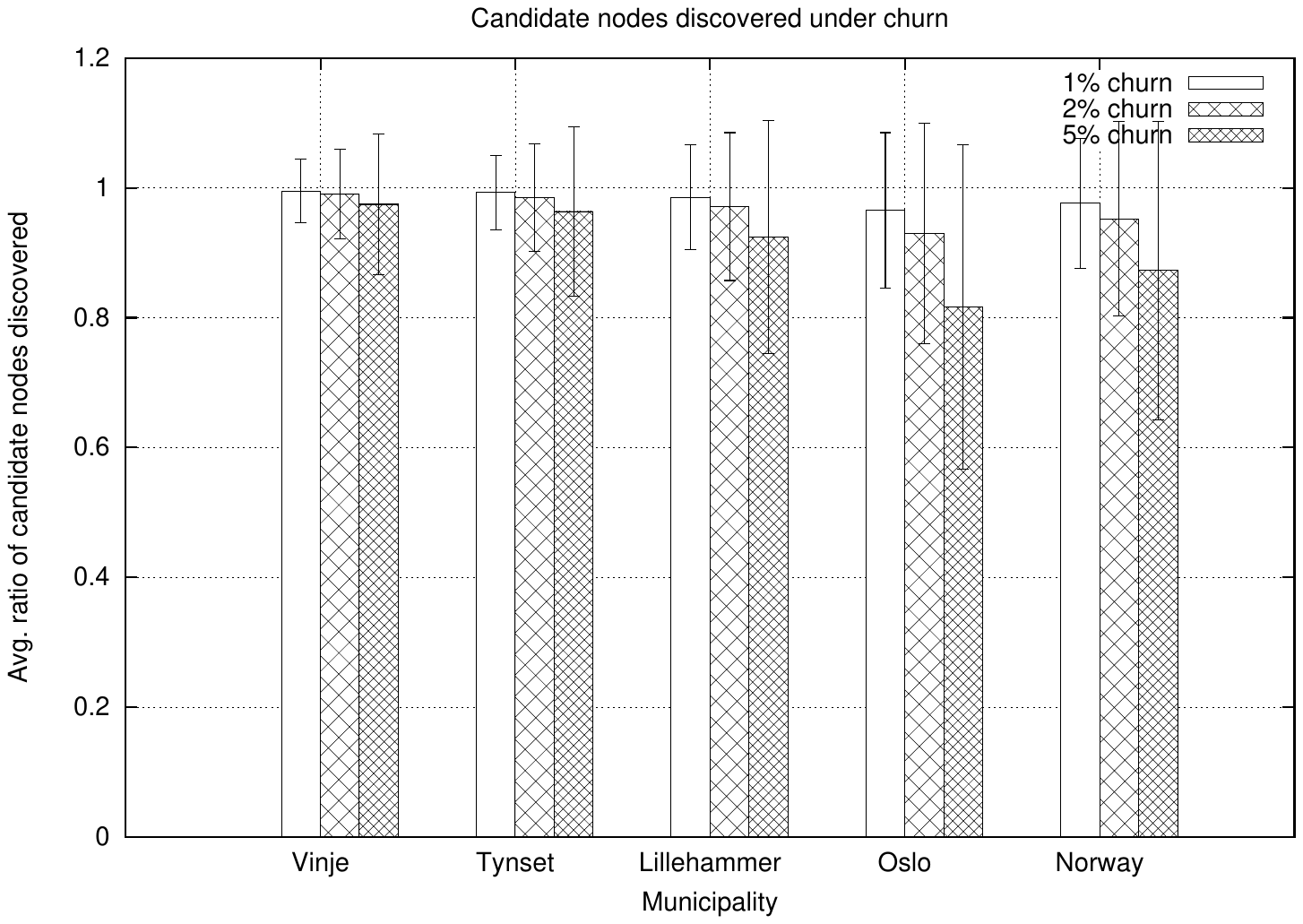}
\caption{Ratio of candidate nodes discovered under continuous churn.}
\label{fig:churnit}
\end{figure}

\begin{figure}
\centering
\includegraphics[trim=2cm 2cm 1cm 2cm, clip=true, width = 0.9\columnwidth]{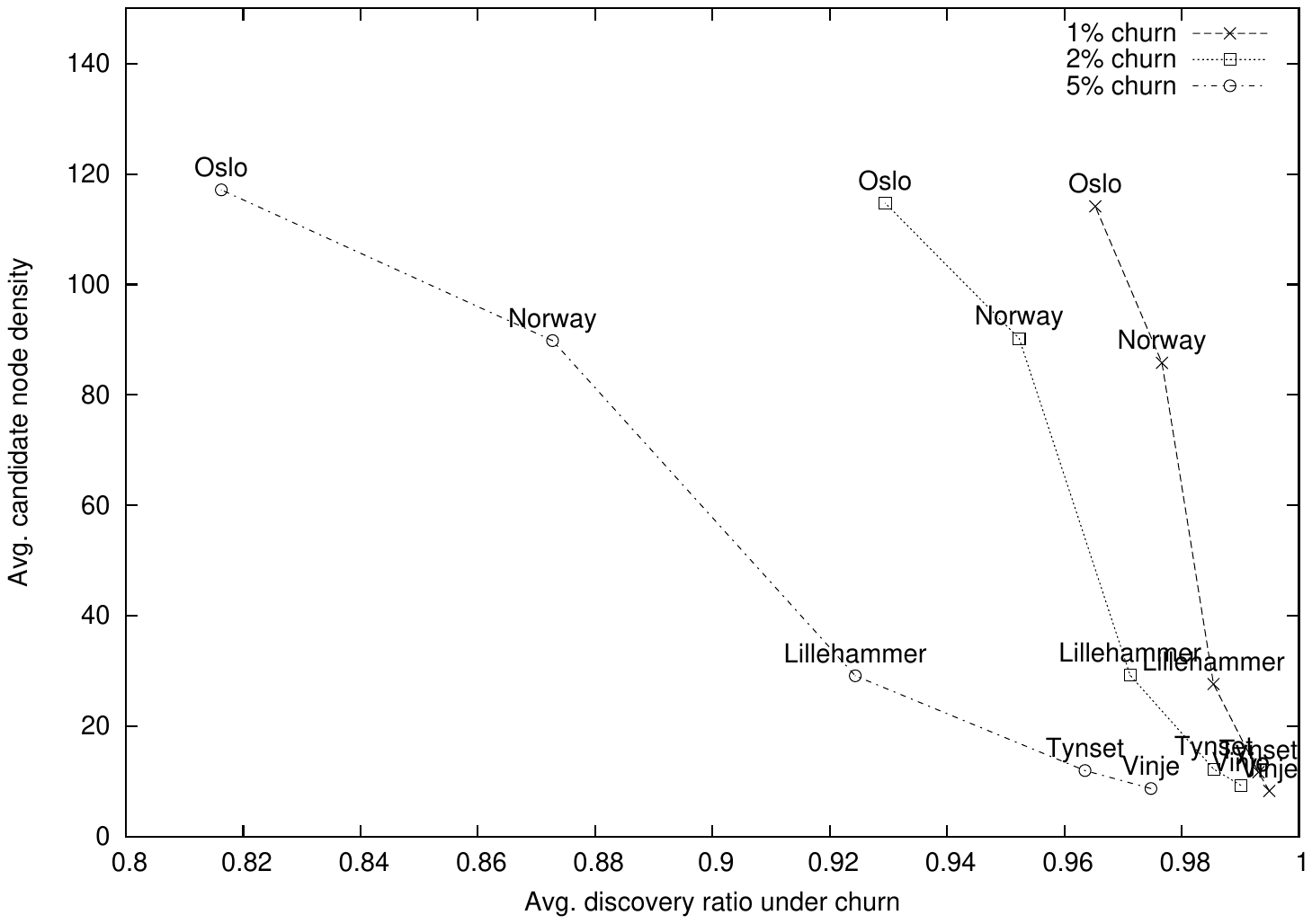}
\caption{Avg.\ discovery ratio under churn by candidate node density.}
\label{fig:churndensity}
\end{figure}

The results are shown in Figure~\ref{fig:churnit}. We can see that for the topologies Vinje and Tynset, the effect of churn is minimal. In Tynset, even with 5\% churn, nodes are able to discover 96.3\% of their candidate nodes on average. In Lillehammer and Oslo, the effect of churn is more noticable, but even in Oslo with 5\% churn the protocol achieves more than a 81\% sustained discovery rate on average. Interestingly, the results for Norway show that the largest topology is more tolerant to churn than the smaller Oslo topology. This indicates that the results are more affected by the average candidate node density of the topology rather than the total number of nodes. Figure~\ref{fig:churndensity} shows the same results plotted by average candidate node density. Here we can clearly see that in our simulations the average discovery ratio increases with the average candidate node density. 

We can see from these results that the protocol performs well in non-uniform topologies with nodes leaving and joining the network, and that it does not break down under heavy churn (5\%). As the protocol is intended to run on devices that are always on it is also likely that the churn in real world applications will be lower.

\section{Related work}
\label{sec:relwork}
Most of the research published so far support the needs of the IEEE 802.22 standard where the challenge is to identify local TV transmitters in the area. This has resulted in numerous publications on detection algorithms and corresponding false alarm and detection probability values. This work is relevant for our work since it will enable us to measure the noise and interference level for a receiver and use locally measured values rather than calculated values. Since we use the Internet for coordination and exchange of radio parameters, the need for a coordinated common control channel is fortunately not required \cite{1264255},  \cite {1542680},  \cite {DARPA_XG} \cite {1443488}.

The idea of using P2P clients at base stations for establishing direct communication via radio using centralized control has been described in a patent claim \cite {Palanki}. This system is aimed at relieving traffic to base stations by establishing direct radio links between users. In another patent claim \cite {Shiquan}, a spectrum manager and a base station controller is used to calculate radio parameters, collect data from the central database and sense the presence of a TV signal. An Internet-based P2P mechanism to locate and distribute data from other clients which is the original part of our paper, is not addressed. 

Gossiping mechanisms have previously been used in cognitive radio networks in \cite{ahmed06}. This approach is different from ours in that it is based on effectively distributing spectral sensing information, not discovering other nodes over the Internet. 

We have previously proposed using P2P for frequency allocation in \cite {5940891} and \cite {5686525}, but the actual discovery protocol has remained future work. 

To the best of our knowledge, there is at this point no other existing P2P protocols that enable discovery of nodes belonging to geographical areas that overlap. 

\section{Conclusion}
\label{sec:conclusion}

In this paper, we have presented a decentralized protocol and architecture for discovering interfering radio devices over the Internet. The protocol has low memory and bandwidth requirements, making it suitable for running on devices with limited hardware. We have shown through simulation that the protocol scales to networks with $2^{20}$ nodes and evaluated its performance in topologies generated from population data from four municipalities in Norway, including Oslo. Our results show that it is possible to implement Internet-scale decentralized protocols that can replace some of the functionality that so far has been implemented in databases. By using decentralized protocols, DSA enabled systems can be deployed in existing networks without the need for new infrastructure. Finally, the P2P protocol has successfully been implemented in C as a prototype in the Open Source Wi-Fi router firmware OpenWRT. In the future, we hope to present experimental results from this platform.

{\footnotesize \bibliographystyle{IEEEtran}

\bibliography{rfc,bib,library}}

\end{document}